\documentclass[a4paper,11pt]{article}
\usepackage{ifpdf}
\usepackage{ifxetex}
\ifpdf
\usepackage[T1]{fontenc}
\usepackage[utf8]{inputenc}
\fi
\usepackage[english]{babel}

\usepackage[
sorting=none,
backend=biber,
doi=false,
isbn=false]{biblatex}

\usepackage{microtype}
\usepackage[left=3.5cm,right=3.5cm,top=4cm,bottom=4cm]{geometry}
\usepackage{fourier}

\usepackage{booktabs}
\usepackage{multirow}
\usepackage{csquotes}
\usepackage{amsmath}
\usepackage{amssymb}
\usepackage{siunitx}
\DeclareSIUnit{\nothing}{\relax}
\usepackage{circuitikz}
\usepackage{cleveref}

\usepackage{authblk}
\usepackage[font=small,labelfont=bf]{caption}
\usepackage{subcaption}
\usepackage{setspace}
\author{Andreas Stöckel}
\author{Aaron R. Voelker}
\author{Chris Eliasmith}
\affil{Centre for Theoretical Neuroscience, University of Waterloo}
\title{Point Neurons with Conductance-Based Synapses in the Neural Engineering Framework}
\date{\today}

\DeclareMathOperator{\diag}{diag}

\begin{document}
\maketitle

\begin{abstract}
The mathematical model underlying the Neural Engineering Framework (NEF) expresses neuronal input as a linear combination of synaptic currents. However, in biology, synapses are not perfect current sources and are thus nonlinear. Detailed synapse models are based on channel conductances instead of currents, which require independent handling of excitatory and inhibitory synapses. This, in particular, significantly affects the influence of inhibitory signals on the neuronal dynamics. In this technical report we first summarize the relevant portions of the NEF and conductance-based synapse models. We then discuss a na\"ive translation between populations of LIF neurons with current- and conductance-based synapses based on an estimation of an average membrane potential. Experiments show that this simple approach works relatively well for feed-forward communication channels, yet performance degrades for NEF networks describing more complex dynamics, such as integration.
\end{abstract}

\section{Introduction}

As a central assumption, the Neural Engineering Framework (NEF, \cite{eliasmith2003neural}) relies on linear interaction of pre-synaptic activity in the form of synaptic currents injected into the neuron membrane. This assumption is readily fulfilled by current-based synapse models. Biological synapses however are not perfect current sources. Instead, currents injected into the cell membrane are caused by changes in receptor channel conductance directly or indirectly gated by neurotransmitters released in the pre-synapse. Consequently, current flow into the membrane is modulated both by pre-synaptic spikes, as well as---due to the nature of ion channels---the difference between the neuron membrane potential and the specific ion species reversal potential. This nonlinear interdependence is modelled by conductance-based synapse models \cite{koch1999biophysics,gerstner2002spiking}.\footnote{Not to be confused with conductance-based neuron models, such as the Hodgkin-Huxley model of action potential generation \cite{hodgkin1952quantitative,koch1999biophysics}.}

For various reasons, conductance-based synapses are a useful addition to the NEF. Foremost, when modelling functional, biologically plausible neural networks, more realistic synapse models pose additional constraints on the network design and reduce degrees of freedom. For example, conductance-based synapses strictly differentiate between excitatory and inhibitory synapses, allowing the network designer to map their parameters to those derived from neurophysiological evidence. Furthermore, some neuromorphic hardware systems (e.g.~Spikey, BrainScaleS; \cite{schemmel2010waferscale,pfeil2013six}) exclusively implement conductance-based synapses, which must be taken into account when assembling networks for these platforms. Finally, nonlinear synaptic interaction has been argued to be an important element of biological neural networks, and as a site for biological computation \cite{koch1999biophysics}. Exploiting these nonlinearities in the NEF increases the computational power per neuron. In the future, conductance-based synapses may thus be utilized to increase the efficiency of certain mathematical operations such as multiplication, requiring fewer neurons at the same precision.

The remainder of this report is structured as follows: \cref{sec:nef} gives a short overview of the relevant equations in the NEF. \Cref{sec:cond_syn} summarizes the biological background of conductance-based synapses and describes the underlying theoretical model. \Cref{sec:nef_cond} elaborates on the actual integration of conductance-based synapses into the NEF and provides an equation highly similar to the original NEF formalism, yet containing additional, voltage-dependent scaling factors. \Cref{sec:ev} provides estimations for these factors based on the expected value for the average membrane potential. \Cref{sec:benchmark} presents two benchmark experiments, followed by concluding remarks in \cref{sec:conclusion}.

\section{Current sources in the Neural Engineering Framework}
\label{sec:nef}

The NEF is a systematic approach to the representation of low-dim\-en\-sio\-nal vectors in populations of spiking neurons and the approximation of arbitrary functions in the connections between these populations \cite{eliasmith2003neural}. As an example, consider two populations of $n$ spiking neurons with all-to-all connectivity from the first (pre-synaptic) to the second (post-synaptic) population. Let $\vec x(t)$ be the value represented by the pre-synaptic population, and $\vec y(t)$ the value represented by the post-synaptic population. Given a function $f$, the NEF allows us to compute connection weights $w_{ij}$ such that the resulting activity in the post-population represents $\vec y(t) \approx f(\vec x(t))$. 

The formalism underlying the NEF assumes that each neuron $j$ receives its input in the form of a current $J(t)$ injected into its (virtual) neuron membrane.\footnote{Note that $J(t)$ (as well as other currents and conductances) should be read with a subscript $j$ in mind -- all equations, unless explicitly stated otherwise, refer to the $j$th neuron in the post-synaptic population.} This current is the sum of a constant bias current $J^\mathrm{bias}$, and a synaptic current $J^\mathrm{syn}(t)$. The latter is calculated from a weighted sum of filtered input spike events: spikes produced by the $i$th pre-synaptic neuron are modelled as a sum of Dirac pulses $a_i(t)$, and a filter $h(t')$ defines the shape of the post-synaptic current in response to a single input spike:
\begin{align}
	J(t)
		&= J^\mathrm{bias} + J^\mathrm{syn}(t) = J^\mathrm{bias} + \sum_{i = 1}^n w_{ij} (a_i \ast h)(t) \,.
 	\label{eqn:nef_j}
\end{align}

Central to the NEF is the idea of factorizing the connection weight matrix $W$ into gain factors $\alpha_j$, decoding vectors $\vec d_i$, and unit-length encoding vectors $\vec e_j$, where $w_{ij} = \alpha_j \langle \vec e_j, \vec d_i \rangle$. Correspondingly, \cref{eqn:nef_j} can be written as
\begin{align}
 	J(t)
		&= J^\mathrm{bias} + J^\mathrm{syn}(t) = J^\mathrm{bias} + \alpha_j \vec e_j^T \Big( \sum_{i = 1}^n \vec d_i (a_i \ast h)(t) \Big)\,.
 	\label{eqn:nef_j_factored}
\end{align}
The decoders $\vec d_i$ linearly project the high-dimensional activity of the pre-synaptic population onto the desired low-dimensional representation $\vec y(t) = f(\vec x(t))$. Multiplication with unit-length encoders $\vec e_j$ corresponds to a projection of this low-dimensional representation onto the preferred direction of each post-synaptic neuron $j$. Correspondingly, the transformed value is again represented as a high-dimensional population coding, allowing to chain an arbitrary number of neuron populations and transformations.

As a side-effect, factorization of the weight matrix $W$ into decoders and encoders simplifies the calculation of $w_{ij}$. Instead of optimizing weights $w_{ij}$ individually, we can randomly select bias currents $J^\mathrm{bias}_j$, gain factors $\alpha_j$, and encoders $\vec e_j$, and calculate decoders $\vec d_j$ from a least squares linear regression
\begin{align}
    Y = D A \Rightarrow D \approx Y A^T \big(A A^T + I \sigma^2 \big)^{-1} \,,
    \label{eqn:nef_decoders}
\end{align}
where the matrix $D$ is composed of the decoding vectors, $A$ is a matrix of neuron responses in the post-synaptic population for a set of sample points, and $Y$ is a matrix of the desired function outputs for these points. The expression $I \sigma^2$ is a regularization term which accounts for uncertainty and noise in the neuron response.

\begin{figure}
    \small
    \centering
    \begin{circuitikz}[american currents]
        \draw(0, 4) to[R, l=$g_{\mathrm{L}}$] (0, 2);
        \draw(0, 2) to[battery1, l=$E_{\mathrm{L}}$] (0, 0);
        \draw(1.8, 0) to[I, i_>=$J^\mathrm{syn}(t)$] (1.8, 4);
        \draw(4, 0) to[I, i_>=$J^\mathrm{bias}$] (4, 4);
        \draw(0, 0) to[short,-*]
             (1.8, 0) to[short,-*] (4, 0);
        \draw(0, 4) to[short,-*]
             (1.8, 4) to[short,-*] (4, 4);
        \draw(4, 0) -- (6, 0) -- (6, 1.85);
        \draw (6, 2.15) -- (6, 4) -- (4, 4);
        \draw (6, 2.15) to[C, l=$C_\mathrm{m}$] (6, 1.85);
        \draw[->] (7.5, 0.5) -- node[right] {$v(t)$} (7.5, 3.5);
    \end{circuitikz}
    \caption{Equivalent circuit diagram of a LIF neuron with current-based synapses (subthreshold behaviour; without spike generation/reset mechanism).}
    \label{fig:circuit_lif_cur}
\end{figure}
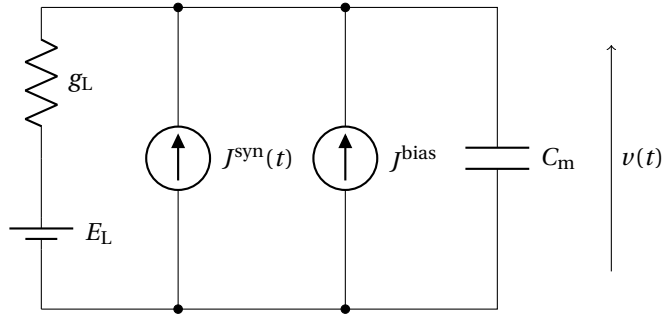
Crucially, the above equations contain the implicit assumption that the input current $J(t)$---and in turn the effect of each individual input spike---is independent of the neuron state at time $t$. Furthermore, all input spikes are treated equally (except for the scaling factor $w_{ij}$), and there is no distinction between excitatory and inhibitory synapses. Having synaptic currents which are independent of the neuron state, is commonly referred to as using \enquote{current-based synapses} \cite{koch1999biophysics,roth2009modeling}.

The subthreshold circuit diagram of a LIF neuron with current-based synapses is depicted in \cref{fig:circuit_lif_cur}. The dynamical system describing this circuit is\footnote{The ratio between the membrane capacitance and the leak channel conductance is the membrane time constant $\tau_\mathrm{RC} = C_\mathrm{m} / g_\mathrm{L}$.}
\begin{align}
	C_\mathrm{m} \frac{\mathrm{d} v(t)}{\mathrm{d}t} = g_\mathrm{L} (E_\mathrm{L} - v(t)) + J^\mathrm{bias} + J^\mathrm{syn}(t) \,,
	\label{eqn:lif_cur}
\end{align}
which has a stable attractor (the equilibrium potential $E_\mathrm{eq}$) at
\begin{align}
	E_\mathrm{eq}(t) = \frac{J^\mathrm{bias} + J^\mathrm{syn}(t)}{g_\mathrm{L}} + E_L \,.
	\label{eqn:lif_cur_eeq}
\end{align}
The synaptic current $J^\mathrm{syn}(t)$ drives the membrane potential $v(t)$ towards more positive values for input spikes received at excitatory synapses ($w_{ij} > 0$) and towards more negative values for spikes received at inhibitory synapses ($w_{ij} < 0$). Importantly, the equilibrium potential $E_\mathrm{eq}$ in \cref{eqn:lif_cur_eeq} can be driven to arbitrary values if the input currents $J$ are large enough. In practice this is not an issue for positive $J$, as the maximum membrane potential is limited by the neuron threshold potential $v_\mathrm{th}$. However, such a natural limitation does not exist for negative currents, and the NEF puts no constraint on the negative current strength, for neither $J^\mathrm{bias}$, nor $J^\mathrm{syn}$. A solution to this problem is to modify the LIF model, such that the membrane potential is clamped to a minimum value\footnote{This is for example implemented in the LIF neuron model in the software neuron simulator included in Nengo, the reference software implementation of the NEF.}, i.e.,~the reset potential $v_\mathrm{reset}$. Note that all experiments in this report are based on LIF neurons with such a minimum membrane potential.

This problem does not occur for conductance-based synapses. Conductance-based synapses are not only biologically motivated, but eliminate the physically implausible, perfect current sources altogether. Unfortunately, and what will be the focus of the remainder of this report, conductance-based synapses introduce a dependency between the neuron membrane potential $v(t)$, and the injected current $J^\mathrm{syn}(t)$. As a result, the post synaptic currents (PSCs) of pre-synaptic spikes no longer combine linearly. Consider two excitatory input spikes which do not trigger an action potential at the receiving neuron. Due to the increased membrane potential $v(t)$, the PSC induced by the second spike is smaller than the PSC induced by the first spike. Furthermore, conductance-based synapses require excitatory and inhibitory synapses to be handled differently. Both issues are not accounted for in the standard NEF \cref{eqn:nef_j,eqn:nef_j_factored}.

\section{Conductance-based synapses}
\label{sec:cond_syn}

\begin{figure}
	\centering
	\begin{subfigure}[b]{0.45\linewidth}
		\centering
		\includegraphics{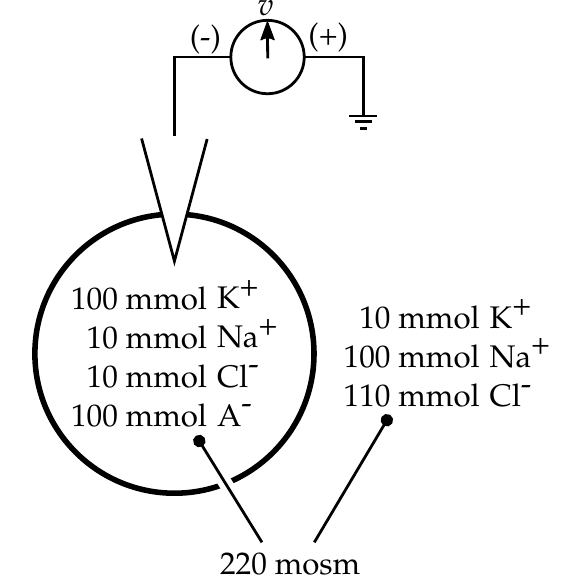}
		\hspace{0.5cm}
		\caption{Membrane without \(\mathrm{K}^+\)-channel}
		\label{fig:neuron_channel_a}
	\end{subfigure}
	\begin{subfigure}[b]{0.45\linewidth}
		\centering
		\includegraphics{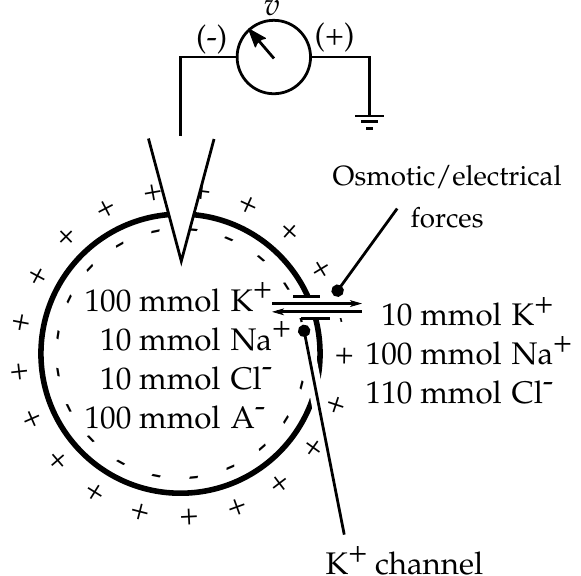}
		\hspace{0.5cm}
		\caption{Added \(\mathrm{K}^+\)-channel}
		\label{fig:neuron_channel_b}
	\end{subfigure}
	\caption{Illustration of the ion channel reversal potential caused by equalling osmotic and electric forces. Although the intra- and extracellular fluids are electrically neutral (a), an open, selectively permeable ion channel causes a change in the membrane potential (b). Illustration adapted from \cite{reichert2000neurobiologie}. See \cite{koch1999biophysics,kandel2012principles} for more details.}
\end{figure}

In biology, changes in the neuron membrane potential are primarily caused by variations in the permeability---or conductance---of selective ion channels embedded into the lipid bilayer cell membrane \cite{koch1999biophysics,kandel2012principles}. As depicted in \cref{fig:neuron_channel_a}, ions are dissolved in different concentrations in the intra- and extracellular fluid. Ion channels provide a way for a specific ion species to follow the osmotic concentration gradient, which in turn causes a charge imbalance and an electrical force countering the osmotic force (\cref{fig:neuron_channel_b}). The charge imbalance itself is measurable as a change in the membrane potential $v(t)$. Once a certain membrane potential is reached, the osmotic and electric forces cancel and the ion flow stops. Further increase/decrease of $v(t)$ reverses the ion flow through a particular channel, giving rise to the name \emph{reversal potential} for this particular $v(t)$.\footnote{Since ion diffusion is a stochastic process, ions will always diffuse in and out of the cell as long as the channel is open. The above claims holds for the average ion flow only.} The exact value of the reversal potential differs between ion channel types. For example, in mammalian cells, the reversal potential for a sodium channel is $E_\mathrm{Na}^+ \approx 67\,\mathrm{mV}$, while for potassium the reversal potential is $E_\mathrm{K}^+ \approx -83\,\mathrm{mV}$. In the resting state, the membrane is slightly permeable for a mixture of ions, resulting in an equilibrium potential\footnote{To clarify the difference between reversal and equilibrium potential: the reversal potential is the potential at which the flow of a \emph{single} ion species reverses direction, the equilibrium potential is the potential at which the sum of \emph{all} currents is zero.} of $E_\mathrm{L} \approx -65\,\mathrm{mV}$ \cite{gerstner2002spiking,gerstner2014neuronal}.

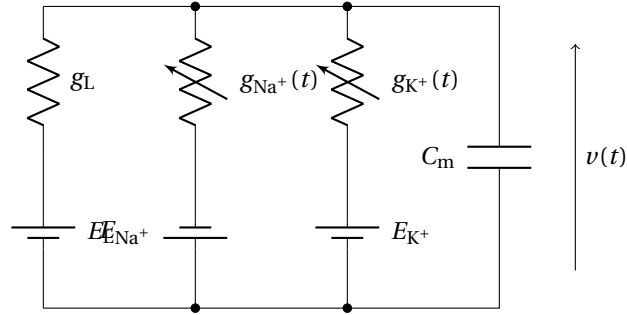
\begin{figure}
	\small
	\centering
	\begin{circuitikz}
		\draw(0, 4) to[R, l=$g_\mathrm{L}$] (0, 2);
		\draw(0, 2) to[battery1, l=$E_\mathrm{L}$] (0, 0);
		\draw(2, 4) to[vR, l=$g_{\mathrm{Na}^+}(t)$] (2, 2);
		\draw(2, 0) to[battery1, l=$E_{\mathrm{Na}^+}$, mirror] (2, 2);
		\draw(4, 4) to[vR, l=$g_{\mathrm{K}^+}(t)$] (4, 2);
		\draw(4, 2) to[battery1, l=$E_{\mathrm{K}^+}$] (4, 0);
		\draw(0, 0) to[short,-*]
		     (2, 0) to[short,-*] (4, 0);
		\draw(0, 4) to[short,-*]
		     (2, 4) to[short,-*] (4, 4);
		\draw(4, 0) -- (6, 0) -- (6, 1.85);
		\draw(6, 2.15) -- (6, 4) -- (4, 4);
		\draw (6, 1.85) to[C, l=$C_\mathrm{m}$] (6, 2.15);
		\draw[->] (7, 0.5) -- node[right] {$v(t)$} (7, 3.5);
	\end{circuitikz}
	\caption{Equivalent circuit diagram of a LIF neuron with conductance-based synapses (without spike generation/reset mechanism), see also \cite{koch1999biophysics}.}
	\label{fig:circuit_lif_cond}
\end{figure}
An equivalent circuit diagram of a cell membrane popularized by Hodgkin and Huxley~\cite{hodgkin1952quantitative} (though in the context of action potential generation, and not synaptic input) with leak, sodium and potassium channel is shown in \cref{fig:circuit_lif_cond}.  
Post-synaptic currents can be modeled as a change of permeabilies in the $\mathrm{Na}^+$ and $\mathrm{K}^+$ ion channel.\footnote{In biology, a large number of channels are embedded into the neuron cell membrane. Individual ion channels are either fully open or fully closed, following a stochastic process. The conductance $g(t)$ can be interpreted as the probability of individual channels to be open.} For example, excitatory input spikes could cause the sodium channels to be open for a certain period of time, driving the membrane potential towards $E_{\mathrm{Na}^+}$. Conversely, inhibitory input spikes might cause potassium channels to open, driving the membrane potential towards $E_{\mathrm{K}^+}$. In the following, we refer to $E_{\mathrm{Na}^+}$ as $E_\mathrm{E}$ (the excitatory reversal potential) and to $E_{\mathrm{K}^+}$ as $E_\mathrm{I}$ (the inhibitory reversal potential). The dynamics for conductance-based synapses in response to incoming spikes $a(t)$ can be described with a linear filter $h(t')$, where the channel conductance $g(t)$ is $g(t) = (a \ast h)(t)$. Possible filters include simple exponential filters, the alpha function, or the difference of exponentials \cite{roth2009modeling}. Of course, in biology, channel dynamics involve the release of a neurotransmitter, diffusion of the transmitter through the synaptic cleft, and a chemical cascade causing ion channels to open as soon as the transmitter reaches the post-synaptic neuron. More detailed models of these highly nonlinear processes exist \cite{roth2009modeling}. Nevertheless, in the remainder of this report we assume the synaptic dynamics to be describable as a linear filter, and in particular focus on normalized exponential low-pass filters with time constant $\tau_\mathrm{syn}$ where
\begin{align*}
	h(t) &= \begin{cases}
		\frac{1}{\tau_\mathrm{syn}} \exp(-t / \tau_\mathrm{syn}) & t \geq 0 \\
		0 & t < 0
	\end{cases} \,, & \text{and} \quad\quad
	\frac{\mathrm{d} g(t)}{\mathrm{d}t} &= -\frac{1}{\tau_{\mathrm{syn}}} \Big( g(t) + \sum_{i = 1}^n w_{ij} a_i(t) \Big) \,.
\end{align*}
Note that while the synaptic dynamics are linear with respect to the input, the post synaptic current is not (see \cref{sec:nef}).

The differential equation describing a LIF point neuron with conductance-based channels is
\begin{align}
	C_\mathrm{m} \frac{\mathrm{d} v(t)}{\mathrm{d}t} = g_\mathrm{L} \cdot (E_\mathrm{L} - v(t)) + g_\mathrm{E}(t) \cdot (E_\mathrm{E} - v(t)) + g_\mathrm{I}(t) \cdot (E_\mathrm{I} - v(t))\,.
	\label{eqn:lif_cond}
\end{align}
The equilibrium potential $E_\mathrm{eq}(t)$ for this model is a weighted average of the individual reversal potentials
\begin{align}
	E_\mathrm{eq}(t) = \frac{g_\mathrm{L} \cdot E_\mathrm{L} + g_\mathrm{E}(t) \cdot E_\mathrm{E} + g_\mathrm{I}(t) \cdot E_\mathrm{I}}{g_\mathrm{L} + g_\mathrm{E}(t) + g_\mathrm{I}(t)} \,.
	\label{eqn:lif_cond_eeq}
\end{align}
This implies that---in contrast to current-based models, which, as noted at the end of the previous section, can provide arbitrarily large input---the membrane potential can never be driven to values outside the voltage range spanned by $E_\mathrm{L}$, $E_\mathrm{E}$, and $E_\mathrm{I}$, or, in biological terms, the ion channel reversal potentials.

\section{Conductance-based synapses in the NEF}
\label{sec:nef_cond}

One of the initial challenges when implementing conductance-based synapses in the NEF is the different handling of excitatory and inhibitory synapses. Spikes arriving at excitatory synapses will influence the excitatory ion channel conductance, whereas spikes arriving at inhibitory synapses influence the inhibitory ion channel conductance.

This issue is virtually nonexistent when adhering to Dale's Principle \cite{parisien2008solving}. Each pre-population of neurons will either project with excitatory or inhibitory connections to the post-population, and influence one of the two channel types only.

In the general case, the weight matrix $W$ (\cref{sec:nef}) can be split into two non-negative matrices $W^+$ and $W^-$. The matrix $W^+$ contains positive, excitatory weights, whereas $W^-$ contains negative, inhibitory weights with all other weights set to zero:
\begin{align*}
	w^+_{ij} &=  \max\{0, w_{ij}\} \,,&
	w^-_{ij} &=  \max\{0, -w_{ij}\} \,.
\end{align*}

This non-negative decomposition of $W$ unfortunately ruins the factorization into encoders and decoders employed by the NEF\@. For two reasons, this is not a real issue. First, decomposition happens after encoders and decoders have already been computed, maintaining the entire NEF pipeline. Second, to increase the computational efficiency during simulation, artificial decoder and encoder matrices $E^\pm$, $D^\pm$ can be calculated, e.g.,~by performing a singular value decomposition of $W^+$ and $W^-$:
\begin{align*}
	  W^\pm &= U S V \,,
		& \vec E^\pm &= U[k] \,,
		& \vec D^\pm &= V[k] \cdot S[k] \,.
\end{align*}
Here, $k$ is the number of non-zero eigenvalues or, alternatively, eigenvalues above a certain threshold, and $M[k]$ refers to the first $k$ columns in the matrix $M$. Under most circumstances, this factorization yields little to no error, since the weight matrix $W$ is an outer product and thus of low rank. Usually, $k$ is equal to the number of dimensions represented by the population plus one (caused by clamping the weights).

As mentioned in the previous section, the channel conductance time-dynamics $g_\mathrm{E}(t)$ and $g_\mathrm{I}(t)$ can be modeled as a linear filter of the input spike trains. Essentially, we can just reuse the post-synaptic current filter $h(t)$ from \cref{eqn:nef_j} and apply it to the individual channel conductances instead. Assuming (as in \cref{sec:nef} and without loss of generality) that all synapses have the same synaptic filter $h(t)$, the conductances $g_\mathrm{E}(t)$ and $g_\mathrm{I}(t)$ for the $j$th neuron are 
\begin{align}
	g_\mathrm{E}(t) &= \sum_{i=1}^n w^+_{ij} \cdot (a_i \ast h)(t)  \,, &
	g_\mathrm{I}(t) &= \sum_{i=1}^n w^-_{ij} \cdot (a_i \ast h)(t)  \,.
	\label{eqn:conductances_filtered}
\end{align}
As per \cref{eqn:lif_cond} the synaptic current $J^\mathrm{syn}(t)$ injected into a single neuron $j$ with conductance-based synapses is
\begin{align}
	J^\mathrm{syn}(t)
		&=   g_\mathrm{E}(t) \cdot \left(E_\mathrm{E} - v(t)\right)
		   + g_\mathrm{I}(t) \cdot \left(E_\mathrm{I} - v(t)\right) \,.
	\label{eqn:j_syn_cond}
\end{align}
Combing \cref{eqn:conductances_filtered,eqn:j_syn_cond} yields
\begin{align}
	J^\mathrm{syn}(t)
		&=   \sum_{i=1}^n \left(
				w^+_{ij} \cdot \left(E_\mathrm{E} - v(t)\right)
			  + w^-_{ij} \cdot \left(E_\mathrm{I} - v(t)\right) \right) \cdot (a_i \ast h)(t) 
	\,.
	\label{eqn:j_syn_cond_2}
\end{align}
This equation is almost equivalent to \cref{eqn:nef_j}, except for the dynamic scaling factors $\alpha_E(t) = E_\mathrm{E}(t) - v(t)$ and $\alpha_I(t) = E_\mathrm{I} - v(t)$. For biologically plausible neuron parameters $v(t) \ll E_\mathrm{E}$, so one might argue that $\alpha_E(t)$ is almost constant. However, for $E_\mathrm{I}$, the difference between the reset or resting potential---if any---is just a few millivolts, causing a strong interdependence between the inhibitory current and $v(t)$ \cite{roth2009modeling}.

\begin{table}[t]
	\newcommand{\sectionhdr}[1]{
		\midrule
		\multicolumn{6}{c}{\emph{#1}} \\
		\midrule
		\multicolumn{2}{c}{\footnotesize\emph{Parameter}} &
		\multicolumn{2}{c}{\footnotesize\emph{Value}} &
		\multicolumn{2}{c}{\footnotesize\emph{Normalized value}} \\
		\cmidrule(r){1-2}\cmidrule(r){3-4}\cmidrule{5-6}
	}
	\caption{Neuron and synapse parameters used throughout this report unless specified otherwise. \enquote{Normalized values} refers to the scaled LIF model commonly used in neuron simulators such as Nengo. Parameters for the conductance-based synapses are scaled according to the linear transition model in \cref{sec:ev}, \cref{eqn:vt_linear}. Synaptic filters are assumed to be normalized to unit area -- whenever this is not the case divide the synaptic weights by $\tau_\mathrm{syn}$ for the denormalized values.}
	\label{tbl:parameters}
	\small
	\centering
	\begin{tabular}{l l r r r r}
		\toprule
		\multicolumn{6}{c}{\textbf{Neuron and synapse parameters}} \\
		\sectionhdr{Neuron parameters}
		Membrane capacitance    & $C_\mathrm{M}$
			& $1.00$ & \si{\nano\farad} 
			& $1.00$ & \\
		Membrane time constant  & $\tau_\mathrm{RC}$
			& $20.00$ & \si{\milli\second}
			& $20.00$ & \si{\milli\second}\\
		Leak conductivity       & $g_\mathrm{L}$
			& $50.00$ & \si{\nano\siemens}
			& $50.00$ & \si{\per\second} \\
		Resting potential       & $E_\mathrm{L}$
			& $-65.00$ & \si{\milli\volt}
			& $0.00$ &\\
		Reset potential         & $v_\mathrm{reset}$
			& $-65.00$ & \si{\milli\volt}
			& $0.00$ & \\
		Threshold potential     & $v_\mathrm{th}$
			& $50.00$ & \si{\milli\volt}
			& $1.00$ & \\
		Refractory period       & $\tau_\mathrm{ref}$
			& $2.00$ & \si{\milli\second}
			& $2.00$ & \si{\milli\second} \\
		\sectionhdr{Current-based synapses}
		Synaptic time constant  & $\tau_\mathrm{syn}$
			& $5.00$ & \si{\milli\second}
			& $5.00$ & \si{\milli\second} \\
		Excitatory weight       & $w^{+}$
			& $35.00$ & \si{\pico\ampere}
			& $2.33$ & $10^{-9}$ \\
		Inhibitory weight       & $w^{-}$
			& $-35.00$ & \si{\pico\ampere} 
			& $-2.33$ & $10^{-9}$ \\
		\sectionhdr{Conductance-based synapses}
		Synaptic time constant  & $\tau_\mathrm{syn}$
			& $5.00$ & \si{\milli\second}
			& $5.00$ & \si{\milli\second} \\
		Excitatory reversal potential & $E_\mathrm{E}$
			& $0.00$ & \si{\milli\volt}
			& $4.33$ & \\
		Inhibitory reversal potential & $E_\mathrm{I}$
			& $-80.00$ & \si{\milli\volt}
			& $-1.00$ & \\
		Excitatory weight       & $w^{+}$
			& $0.61$ & \si{\nano\siemens}
			& $40.60$ & $10^{-9}$ \\
		Inhibitory weight       & $w^{-}$
			& $1.56$ & \si{\nano\siemens}
			& $103.70$ & $10^{-9}$ \\
		\bottomrule
		\end{tabular}
\end{table}

\begin{figure}
	\begin{subfigure}[t]{\linewidth}
		\centering
		\includegraphics{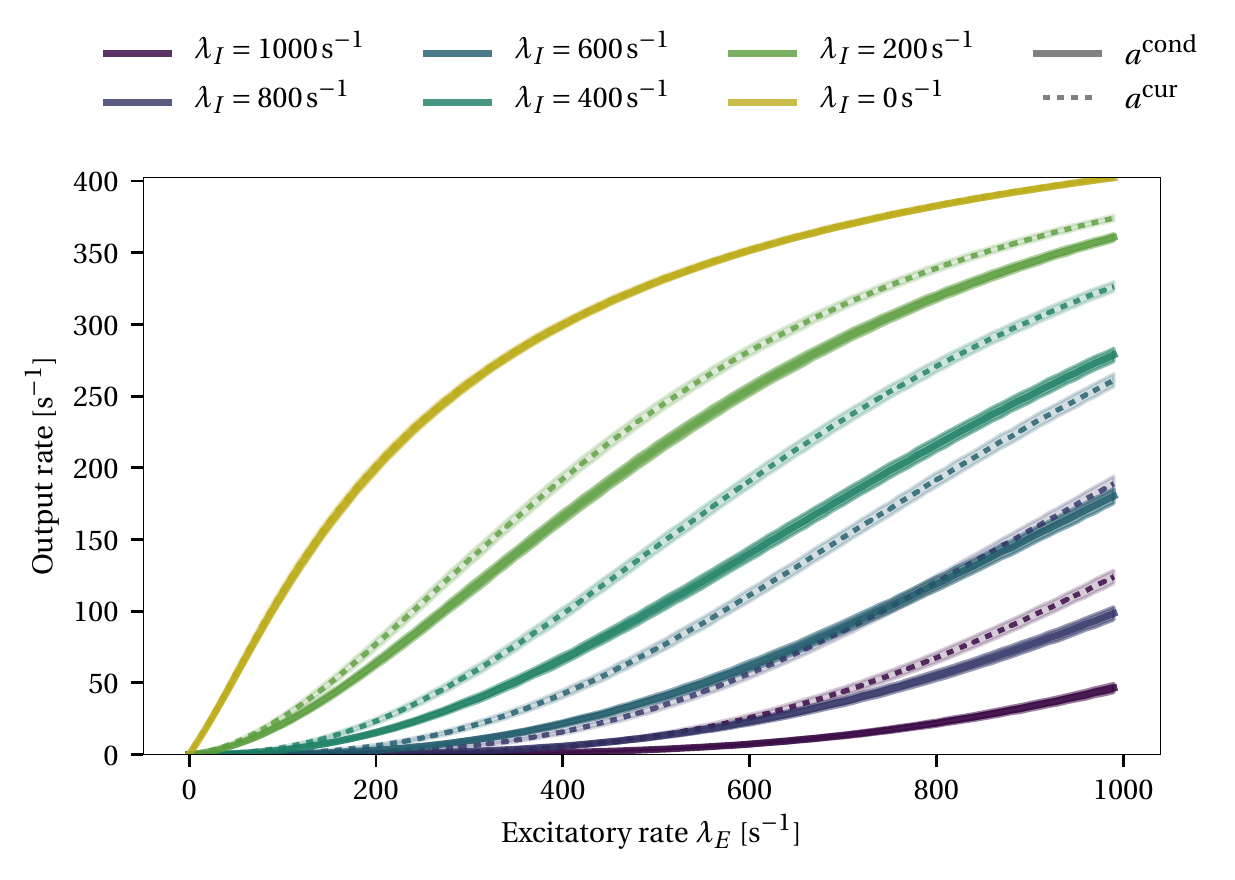}
		\caption{Response curves}
		\label{fig:lif_cur_vs_cond_jbias_rate}
	\end{subfigure}
	\begin{subfigure}[t]{\linewidth}
		\centering
		\includegraphics{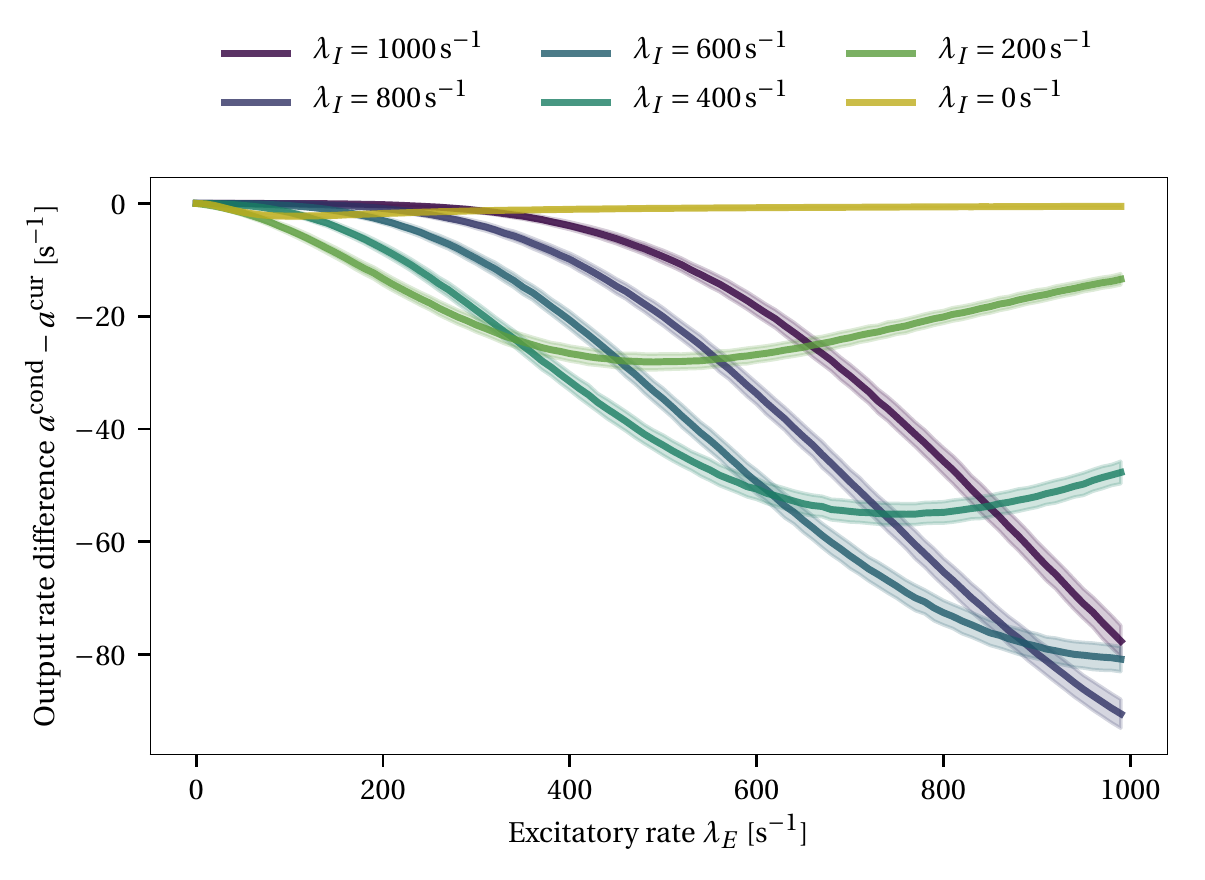}
		\caption{Response curve difference}
		\label{fig:lif_cur_vs_cond_jbias_diff}
	\end{subfigure}
	\caption{Neuron response curves $a^\mathrm{cur}(\lambda_\mathrm{E})$, $a^\mathrm{cond}(\lambda_\mathrm{E})$ of a LIF neuron with current-based or conductance-based synapses under varying inhibitory bias currents $J^\mathrm{bias}$. The excitatory input spike train is Poisson-distributed with rate $\lambda_\mathrm{E}$ over $10\,\mathrm{s}$. Lines represent the mean over a thousand trials, shaded areas the standard deviation.}
	\label{fig:lif_cur_vs_cond_jbias}
\end{figure}
\begin{figure}
	\begin{subfigure}[t]{\linewidth}
		\centering
		\includegraphics{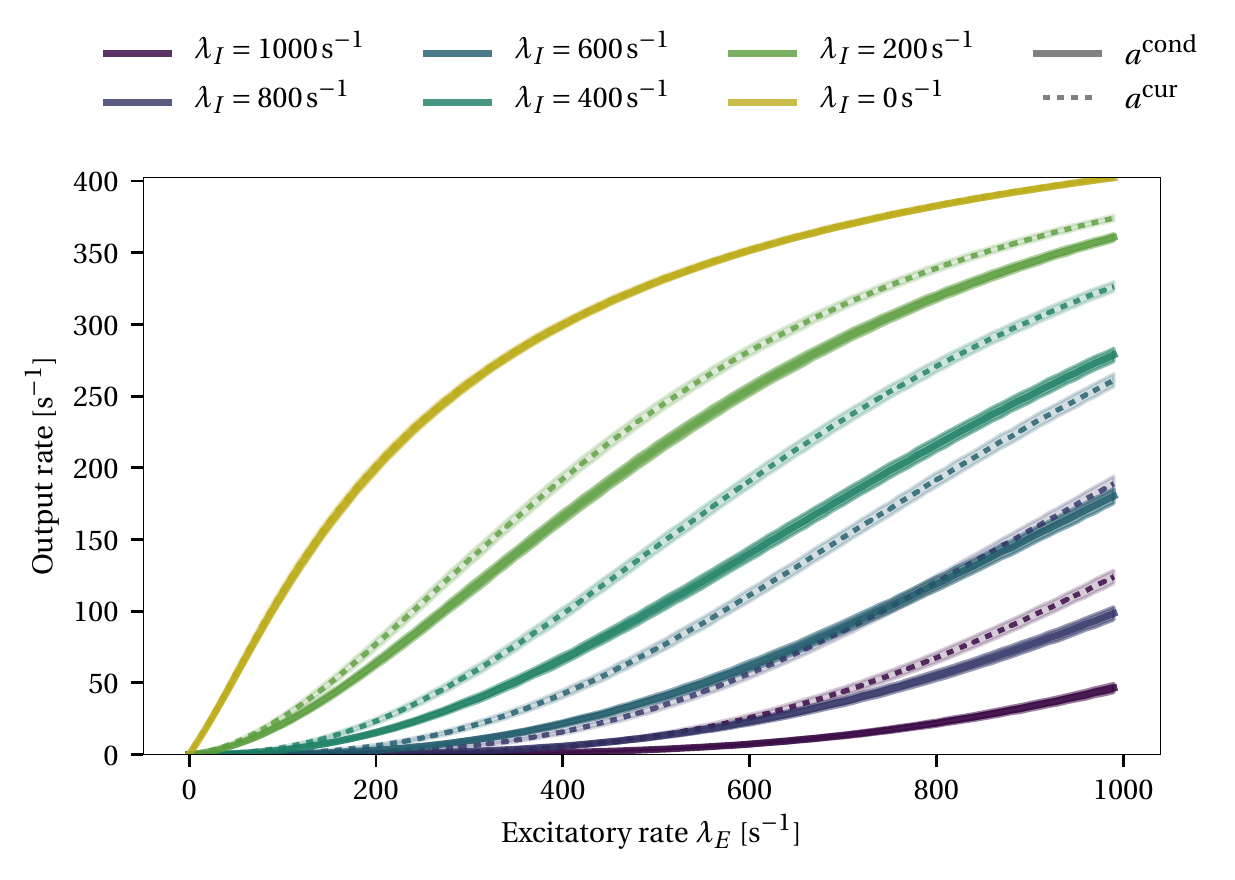}
		\caption{Response curve}
		\label{fig:lif_cur_vs_cond_e_vs_i_rate}
	\end{subfigure}
	\begin{subfigure}[t]{\linewidth}
		\centering
		\includegraphics{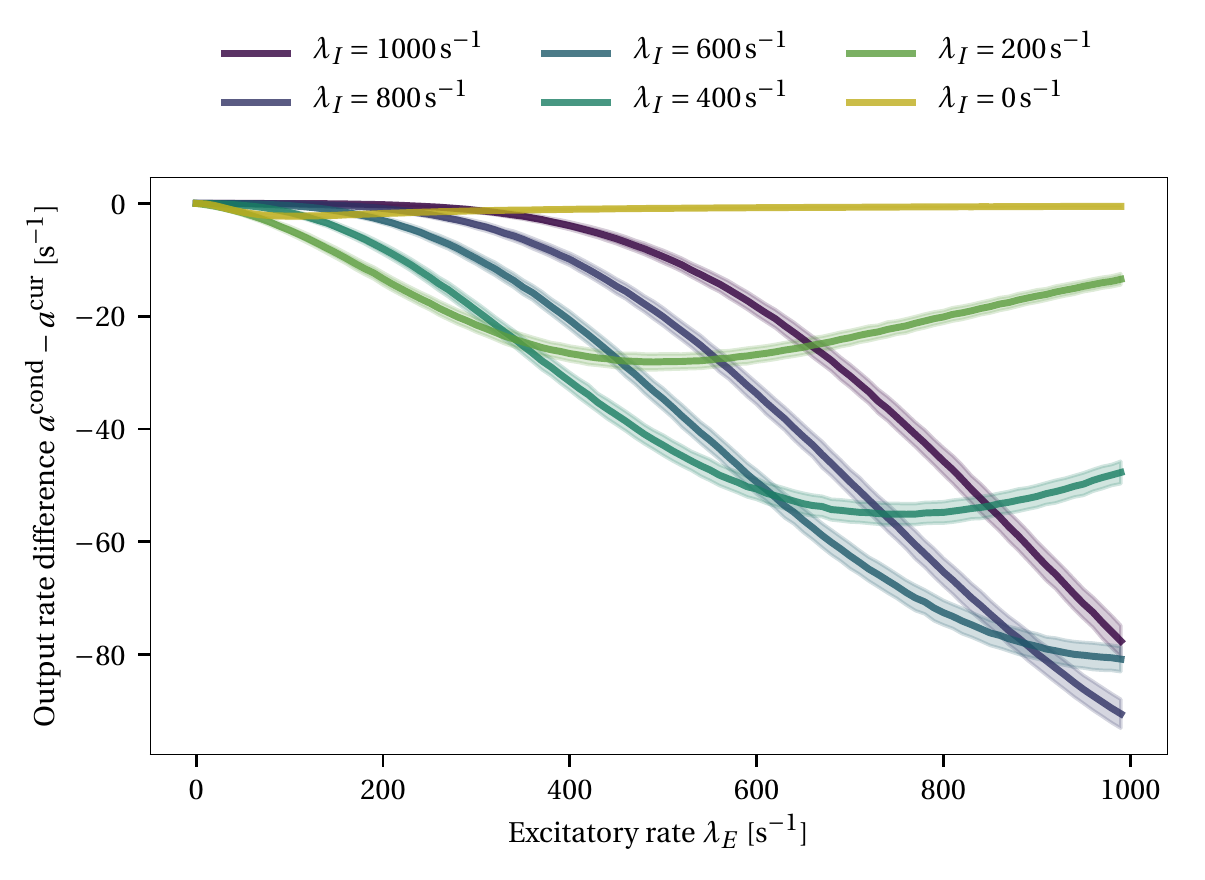}
		\caption{Response curve difference}
		\label{fig:lif_cur_vs_cond_e_vs_i_diff}
	\end{subfigure}
	\caption{Neuron response curves $a^\mathrm{cur}(\lambda_\mathrm{E})$, $a^\mathrm{cond}(\lambda_\mathrm{E})$ of a LIF neuron with current-based or conductance-based synapses for an input spike train arriving at an inhibitory synapse. Both, the excitatory and the inhibitory input spike trains are Poisson-distributed with rates $\lambda_\mathrm{E}$, $\lambda_\mathrm{I}$ over $10\,\mathrm{s}$. Lines represent the mean over a thousand trials, shaded areas the standard deviation.}
	\label{fig:lif_cur_vs_cond_e_vs_i}
\end{figure}
\begin{figure}
	\begin{subfigure}[t]{\linewidth}
		\centering
		\includegraphics{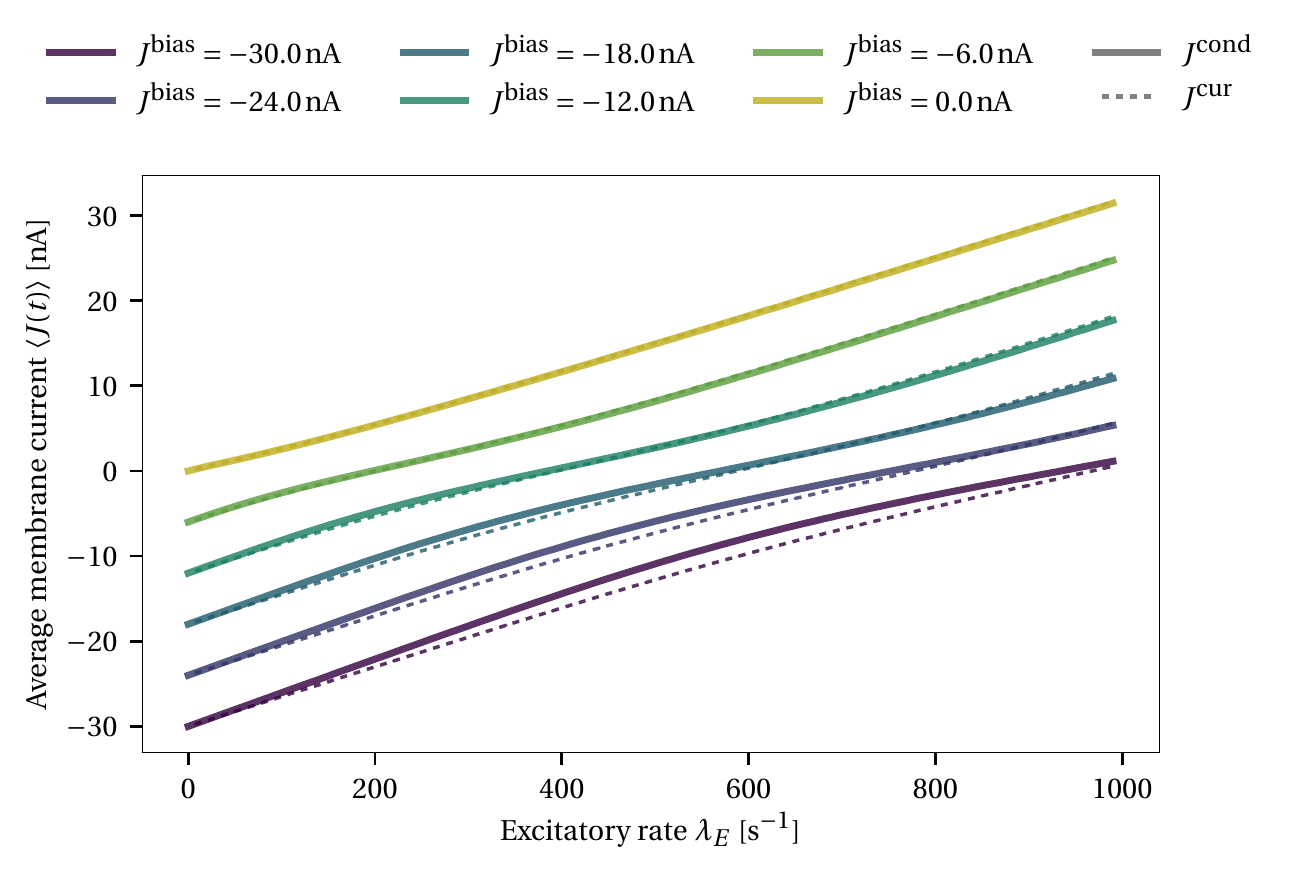}
		\caption{Average membrane current for inhibitory $J^\mathrm{bias}$}
		\label{fig:lif_cur_vs_cond_jbias_cur}
	\end{subfigure}
	\begin{subfigure}[t]{\linewidth}
		\centering
		\includegraphics{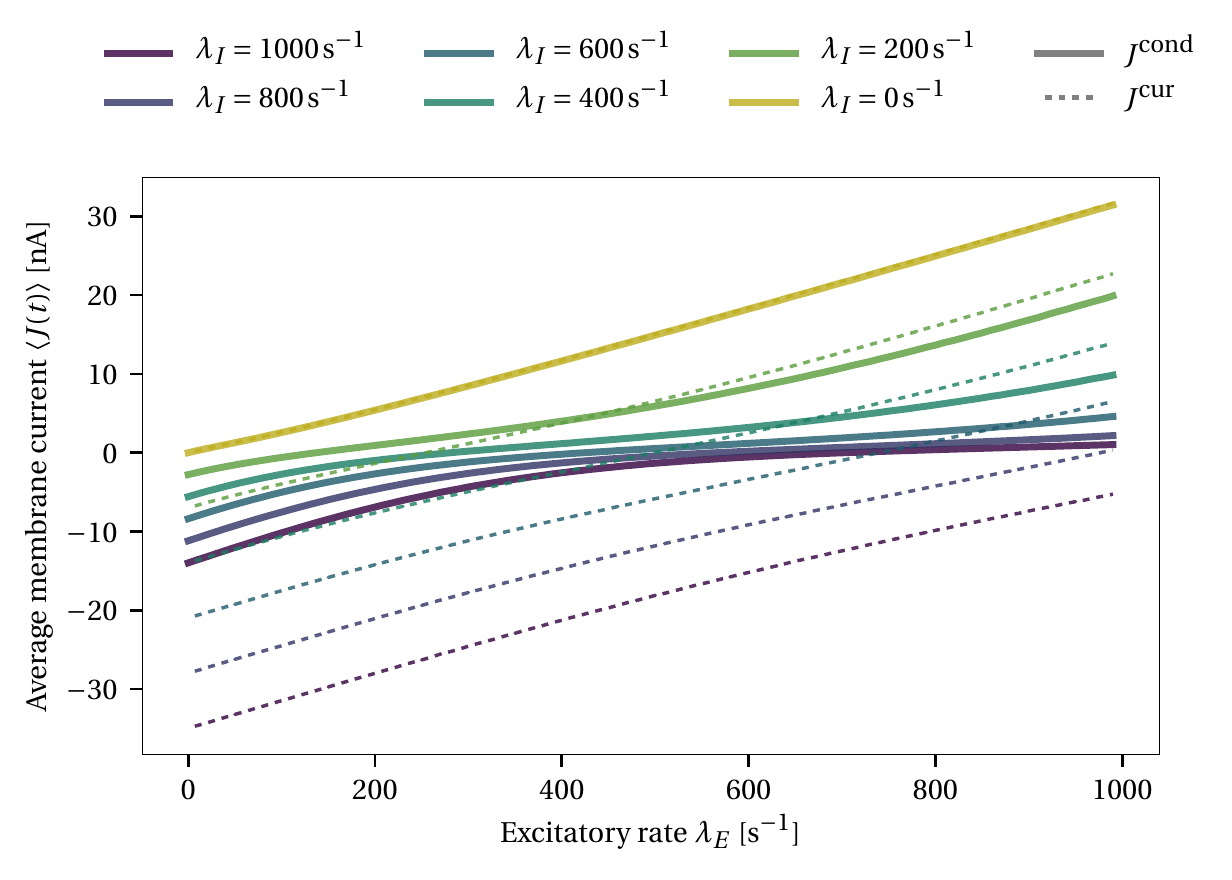}
		\caption{Average membrane current for inhibitory input spikes.}
		\label{fig:lif_cur_vs_cond_e_vs_i_cur}
	\end{subfigure}
	\caption{Average membrane current $J$ of a LIF neuron with conductance-based synapses for an input spike train arriving at an inhibitory synapse. Both, the excitatory and the inhibitory input spike trains are Poisson-distributed with rates $\lambda_\mathrm{E}$, $\lambda_\mathrm{I}$ over $10\,\mathrm{s}$. Lines represent the mean over a thousand trials, shaded areas the standard deviation.}
	\label{fig:lif_cur_vs_cond_cur}
\end{figure}
These potential deviations for inhibitory synapses are supported by the experiments presented in \cref{fig:lif_cur_vs_cond_jbias,fig:lif_cur_vs_cond_e_vs_i}. In each individual experiment, a single LIF neuron with either current or conductance-based synapses is simulated (see \cref{tbl:parameters}). Parameters for the conductance-based synapses are matched to the current-based synapses with the linear transition model discussed in \cref{sec:ev}, \cref{eqn:vt_linear}, so the neuron should behave similarly in both cases. The neuron receives an excitatory input spike train with Poisson statistics and rate $\lambda_\mathrm{E}$. In \cref{fig:lif_cur_vs_cond_jbias} an inhibitory bias current $J^\mathrm{bias}$ is injected into the neuron, whereas in \cref{fig:lif_cur_vs_cond_e_vs_i} the neuron receives spikes at an inhibitory synapse with rate $\lambda_\mathrm{I}$. The inhibitory currents/rates were matched such that the average neuron rate surpasses zero for approximately the same excitatory input (the $x$-intercepts are in similar locations). Strikingly, for excitatory input only ($J^\mathrm{bias} = 0$, $\lambda_\mathrm{I} = 0$), there are only minor deviations between the neuron response curves. For an inhibitory $J^\mathrm{bias}$, the neuron with conductance-based synapses has a slightly smaller output rate than the neuron with current-based synapses, though this difference converges to zero for high output rates. However, if the inhibitory bias current is replaced by an inhibitory synaptic input, the deviations from the neuron with current-based synapses are much stronger: the output rate is significantly reduced in the presence of inhibitory synaptic input, especially for large input spike rates. \Cref{fig:lif_cur_vs_cond_cur} shows the average current flowing into the membrane in the above experiments. For constant inhibitory bias current, the current increases almost linearly with the excitatory spike rate, whereas inhibitory spikes cause a saturation of the input current. This non-linearity is likely caused by inhibitory spikes injecting a larger negative current in the presence of excitatory input due to the increased average membrane potential.

\section{Average membrane potential estimation}
\label{sec:ev}

There are at least two possible approaches regarding the integration of conductance-based synapses into the NEF\@. A \enquote{natural} approach would be to measure the actual neuron response curves $A$ as required for the calculation of decoders in \cref{eqn:nef_decoders}. However, the neuron response is no longer defined in terms of a current $J$ but as a two-dimensional function of $g_\mathrm{E}$ and $g_\mathrm{I}$, which slightly complicates matters, and is not the focus of this report.

Alternatively, we can try to transform a network of neurons with current-based synapses into networks of neurons with conductance-based synapses. Again, this assumes that the switch to conductance-based synapses does not have a strong effect on the neuronal dynamics. For this approach, we calculate weights $W$ as if the network was still using current-based synapses, but rescale the decomposed $W^+$, $W^-$ linearly to account for $\alpha_E(t)$ and $\alpha_I(t)$. Instead of treating these factors as functions time, we estimate expected values $\langle \alpha_E(t) \rangle_t$ and $\langle \alpha_I(t) \rangle_t$. This can be achieved by replacing $v(t)$ with $\langle v(t) \rangle_t$. Clearly, a \enquote{one size fits all} average membrane potential does not exist. However, in the NEF neurons are often in either a mode of (semi) regular spiking or a do not spike at all for longer periods of time.

\begin{figure}
	\centering
	\includegraphics{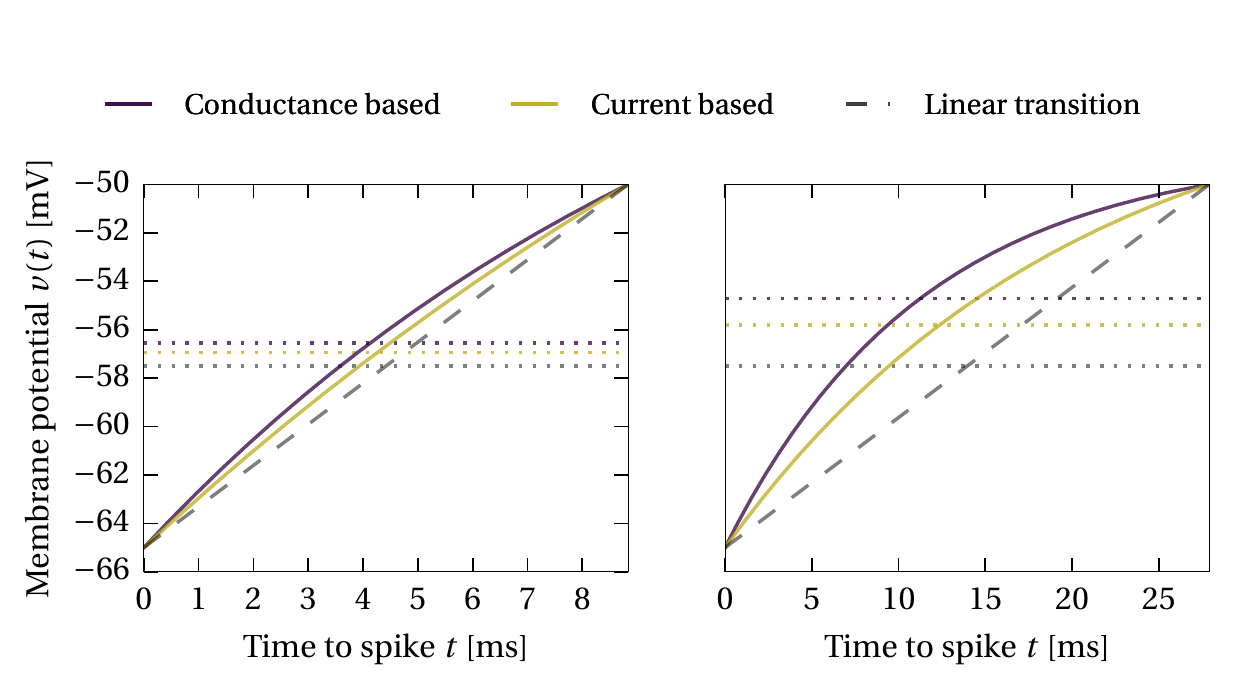}
	\caption{Idealized membrane potential traces. Comparison of the membrane potential transition from $v_\mathrm{reset} = -65\,\mathrm{mV}$ to $v_\mathrm{th} = -50\,\mathrm{mV}$ for a LIF neuron with conductance-based synapses, a LIF neuron with current-based synapses, and a simple linear model. The left and right graphs show different transition times ($9\,\mathrm{ms}$ and $27\,\mathrm{ms}$). Conductances/currents were assumed to be constant. Dotted lines show the average for each of the three models.}
	\label{fig:neuron_transitions_cond_vs_cur}
\end{figure}
A na\"ive estimate for $\langle v(t) \rangle_t$ in the spiking mode is to assume a linear transition between $v_\mathrm{reset}$ and $v_\mathrm{th}$. In this case, the average membrane potential is
\begin{align}
	\langle v(t) \rangle_t &= \frac{v_\mathrm{reset} + v_\mathrm{th}}{2} \,. & \text{(Linear transition)}
	\label{eqn:vt_linear}
\end{align}
However, the membrane potential transition for a neuron is not linear. The force driving the membrane potential towards $E_L$ gets larger with higher $v(t)$, whereas the force driving towards $E_E$ gets smaller (for conductance-based neurons). Correspondingly, and as illustrated in \cref{fig:neuron_transitions_cond_vs_cur}, $v(t)$ progresses more slowly through high voltage regimes, causing the actual $\langle v(t) \rangle_t$ to be larger than the prediction made by the linear model.

Under the assumption of fast enough regular input spikes we can derive a better approximation for the expected value. For high input rates $g_E(t)$ and $g_I(t)$ are approximately constant. Specifically, for an infinite input spike train $a(t)$ with constant inter-spike interval $\Delta t$ and a normalized exponential low pass filter $h(t')$ with time constant $\tau$, it holds (see \cref{app:avg_filter_value} for the derivation)
\begin{align*}
	\langle g(t) \rangle_t = \langle (a \ast h)(t) \rangle_t
	&= \frac{1}{\Delta t} \,.
\end{align*}
For constant conductance, \cref{eqn:lif_cond} can be written as
\begin{align*}
    C_\mathrm{m} \frac{\mathrm{d} v(t)}{\mathrm{d}t} &=
        g_\mathrm{L} \cdot (e_\mathrm{L} - v(t)) + g_\mathrm{E} \cdot (e_\mathrm{E} - v(t)) + g_\mathrm{I} \cdot (e_\mathrm{I} - v(t)) + J^\mathrm{bias} \\
        &= - v(t) \cdot (g_\mathrm{L} + g_\mathrm{E} + g_\mathrm{I}) + g_\mathrm{L} e_\mathrm{L}  + g_\mathrm{E} e_\mathrm{E} + g_\mathrm{I} e_\mathrm{I} + J^\mathrm{bias}\,,
\end{align*}
which is brought into a fairly general form by substituting
\begin{align*}
    \lambda &= \frac{g_\mathrm{L} + g_\mathrm{E} + g_\mathrm{I}}{C_\mathrm{m}} \,,\\
    \mu     &= \frac{e_\mathrm{L} g_\mathrm{L} + e_\mathrm{E} g_\mathrm{E} + e_\mathrm{I} g_\mathrm{I} + J^\mathrm{bias}}{C_\mathrm{m}} \,.
\end{align*}
The closed-form solution for an initial value of $v(0) = v_0$ is
\begin{align*}
    \frac{\mathrm{d} v(t)}{\mathrm{d}t}
        &= - \lambda v(t) + \mu
	\,, &
	v(t) &= \left( v_0 - \frac{\mu}\lambda \right) \cdot e^{-\lambda t} + \frac{\mu}\lambda
	\,.
\end{align*}
If the synaptic current suffices to generate output spikes, the time-to-spike $t_\mathrm{th}$ after which the threshold potential $v_\mathrm{th}$ is reached is given as
\begin{align}
	t_\mathrm{th} = - \frac{1}{\lambda} \log \left( \frac{v_\mathrm{th} - \frac{\mu}{\lambda}}{v_0 - \frac{\mu}{\lambda}} \right) \,.
\end{align}
The average membrane potential in the period between two spikes can now be calculated by setting $v_0 = v_\mathrm{reset}$, integrating over the above equation, and dividing by $t_\mathrm{th}$:
\begin{align*}
	\langle v(t) \rangle_t
		&= \frac{1}{t_\mathrm{th}} \int_0^{t_\mathrm{th}} v(t) \,\mathrm{d}t
		= \frac{v_0 - \frac{\mu}{\lambda}}{\lambda t_\mathrm{th}} \left(1 - e^{-\lambda t_\mathrm{th}} \right) + \frac{\mu}\lambda
\end{align*}
For $g_\mathrm{E} \to \infty$ the expected value $\langle v(t) \rangle_t$ is
\begin{align}
	\langle v(t) \rangle_t &=
	\frac{
		\log\left(\frac{E_\mathrm{E} - v_\mathrm{th}}{E_\mathrm{E} - E_\mathrm{L}}\right) \cdot E_\mathrm{E} - E_\mathrm{L} + v_\mathrm{th}
	}{\log\left(\frac{E_\mathrm{E} - v_\mathrm{th}}{E_\mathrm{E} - E_\mathrm{L}}\right)} \,. & \text{(Conductance-based)}
	\label{eqn:vt_conductance}
\end{align}
Explicitly, these calculations do not include the refractory period $\tau_\mathrm{ref}$. The LIF neuron model assumes all currents flowing into the cell membrane to be zero during the refractory state. As a consequence, conductance-based synapses do not change the behavior of the model during the refractory period, so the average membrane potential calculation should be independent of $\tau_\mathrm{ref}$.

To give a sense for the difference between the two estimates discussed so far: for the neuron parameters listed in \cref{tbl:parameters} the linear transition model \cref{eqn:vt_linear} estimates an average value of \SI{-57.5}{\milli\volt}, whereas the conductance-based transition model estimates a slightly higher value of \SI{-57.17}{\milli\volt} for an infinitely strong excitatory input.

\begin{figure}
	\begin{subfigure}[t]{\linewidth}
		\centering
		\includegraphics{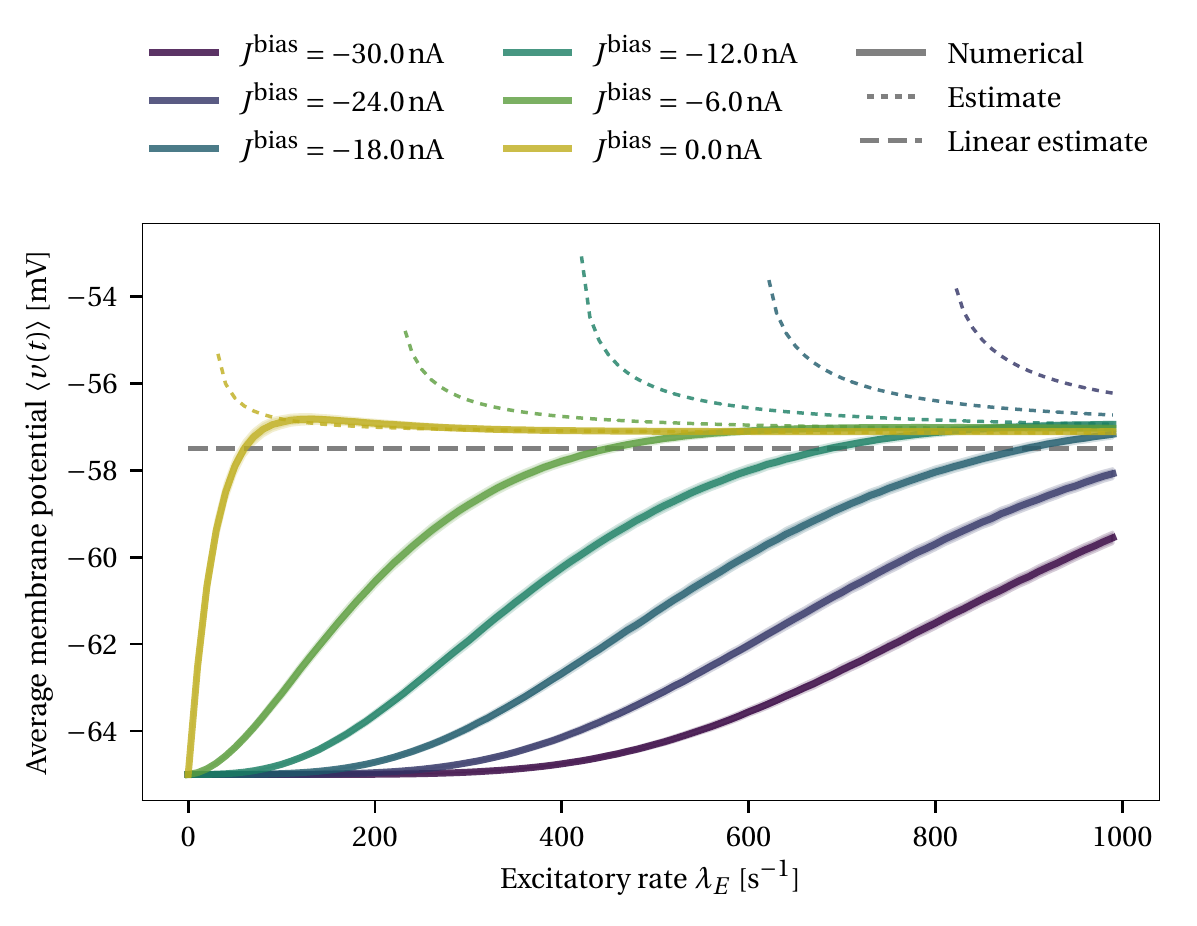}
		\caption{Inhibitory $J^\mathrm{bias}$}
		\label{fig:lif_cur_vs_cond_jbias_volt}
	\end{subfigure}
	\begin{subfigure}[t]{\linewidth}
		\centering
		\includegraphics{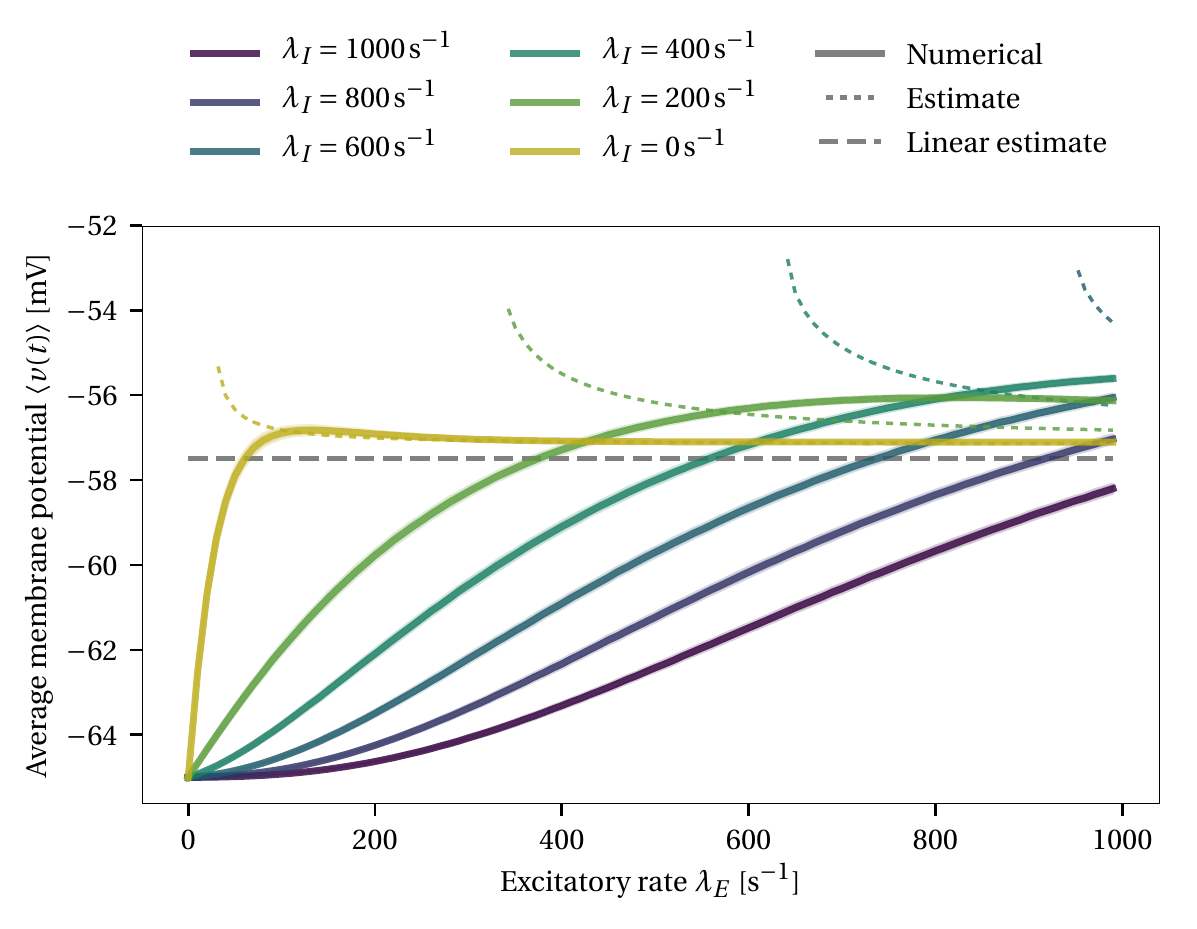}
		\caption{Inhibitory input spikes}
		\label{fig:lif_cur_vs_cond_e_vs_i_volt}
	\end{subfigure}
	\caption{Estimate and measured average membrane potentials for varying excitatory spike rates $\lambda_{E}$ and either a constant bias current $J^\mathrm{bias}$ or an inhibitory spike train with rate $\lambda_{I}$. Lines represent the mean over a thousand trials, shaded areas the standard deviation.}
	\label{fig:lif_cur_vs_cond_volt}
\end{figure}
\Cref{fig:lif_cur_vs_cond_volt} shows numerical measurements of the average membrane potential for the experimental setup presented in the previous section. For a constant bias current $J^\mathrm{bias}$ and no inhibitory spikes, the membrane potential converges to the estimate from \cref{eqn:vt_conductance} as $\lambda_\mathrm{E}$ increases. In the presence of inhibitory input spikes, $\langle v(t) \rangle_t$ is consistently larger than in the $J^\mathrm{bias}$ experiment and reaches values clearly above the estimate. Unintuitively, and despite output-rate deviations being higher (\cref{fig:lif_cur_vs_cond_jbias,fig:lif_cur_vs_cond_e_vs_i}), the mean error between the estimated average membrane potential and the numerically measured value is consistently smaller when using inhibitory spikes instead of a bias current $J^\mathrm{bias}$. This is especially true for high $\lambda^\mathrm{I}$ in comparison to high $J^\mathrm{bias}$ (i.e.~\SI{5.5}{\milli\volt} RMSE for $\lambda^\mathrm{I} = \SI{1000}{\per\second}$, \SI{6.5}{\milli\volt} RMSE for $J^\mathrm{bias} = \SI{-30}{\nano\ampere}$). Consequently, it might be worthwhile to investigate the effect of not using an external current source $J^\mathrm{bias}$ in the NEF altogether without negatively affecting the variety of tuning curves. The next section contains a simple technique achieving exactly this by decoding $J^\mathrm{bias}$ from the pre-population.

\section{Elimination of the bias current}
\label{sec:bias}

As pointed out in \cref{eqn:nef_j_factored}, the NEF generally defines the current injected into the membrane $J(t)$ of a neuron $j$ as $J(t) = J^\mathrm{bias} + J^\mathrm{syn}(t)$. For a population of $n$ neurons, this can be written in matrix notation
\begin{align*}
	\vec J(t) &= \vec J^\mathrm{bias} + \diag(\vec \alpha) \cdot E D \cdot (\vec a \ast h)(t) =  \vec J^\mathrm{bias} + W \cdot (\vec a \ast h)(t)  \,,
\end{align*}
where $\vec J^\mathrm{bias}$ and $\vec \alpha$ are $n$-dimensional vectors, $E$ is the $n \times d$ encoding matrix ($d$ being the dimensionality of the vector represented by the population), $D$ is $d \times m$ decoding matrix ($m$ being the size of the pre-population), $W$ is the $n \times m$ weight matrix, and $(\vec a \ast h)(t)$ is the filtered, $m$ dimensional activity of the pre-population. We can approximate this equation as
\begin{align*}
	\vec J(t) &\approx \left(D' + \diag( \vec \alpha ) \cdot E D \right) \cdot (\vec a \ast h)(t) = W' \cdot (\vec a \ast h)(t)\,,
\end{align*}
where $D'$ is a $n \times m$ matrix decoding the vector $\vec J^\mathrm{bias}$ from the pre-population. Analogously to the $L_2$ approximation in \cref{eqn:nef_decoders}, we can approximate $D'$ as
\begin{align*}
	Y' = \big( \underbrace{\vec J^\mathrm{bias}, \ldots, \vec J^\mathrm{bias}}_{m \text{ times}} \big) = D' A \Rightarrow D' = Y' A^T \left(A A^T + I \sigma^2\right)^{-1} \,,
\end{align*}
where $Y'$ is the $n \times N$ dimensional target matrix (with $N$ being the number of evaluation points), and $A$ the $n \times N$ neuron response matrix. Using the adapted weight matrix $W' = D' + W$ instead of $J^\mathrm{bias}$ should not significantly alter the behaviour of the system. As discussed in \cref{sec:nef_cond}, we can split $W'$ into excitatory weights $W'^+$ and inhibitory weights $W'^-$  for use with conductance-based synapses.

\section{Benchmark experiments}
\label{sec:benchmark}

In this section we present two benchmark networks: a feed-forward communication channel, and an integrator. The first network tests how well one can decode a function from a conductance-based synapse population with respect to pure feed-forward processing and a constant stream of input. The second network assess how well dynamical systems are realized. We then analyze a potential cause of unexpected behaviour in the recurrent network by empirically measuring neuron tuning curves.

\begin{figure}[p]
	\centering
	\begin{subfigure}[t]{\linewidth}
		\centering
		\includegraphics{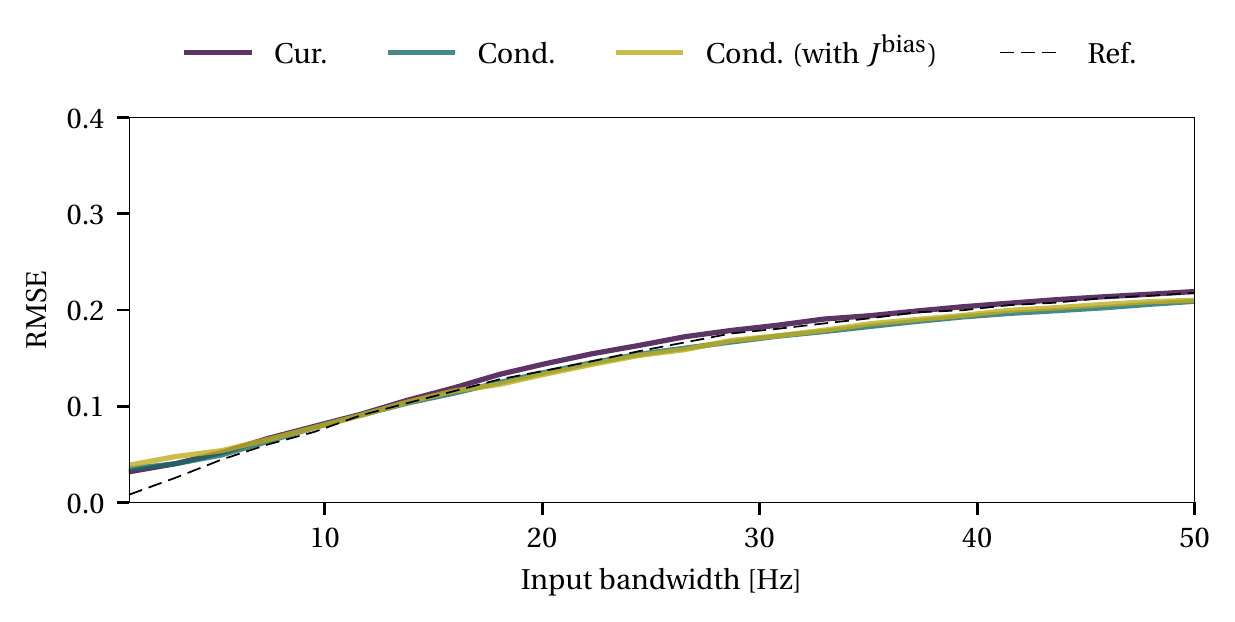}
		\caption{$n = 2$}
	\end{subfigure}
	\begin{subfigure}[t]{\linewidth}
		\centering
		\includegraphics{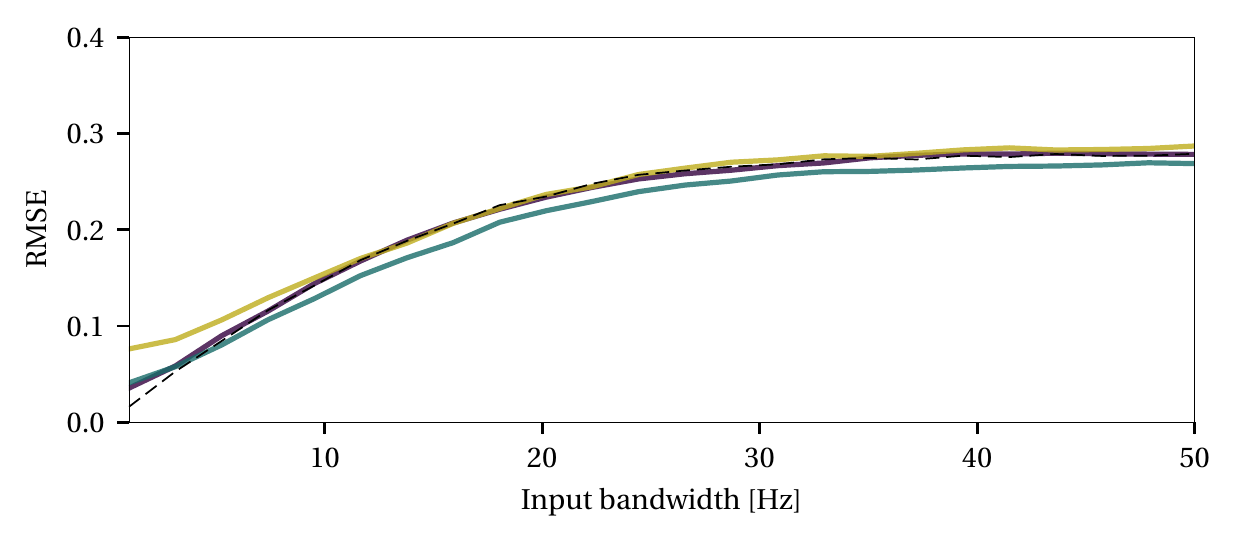}
		\caption{$n = 4$}
	\end{subfigure}
	\begin{subfigure}[t]{\linewidth}
		\centering
		\includegraphics{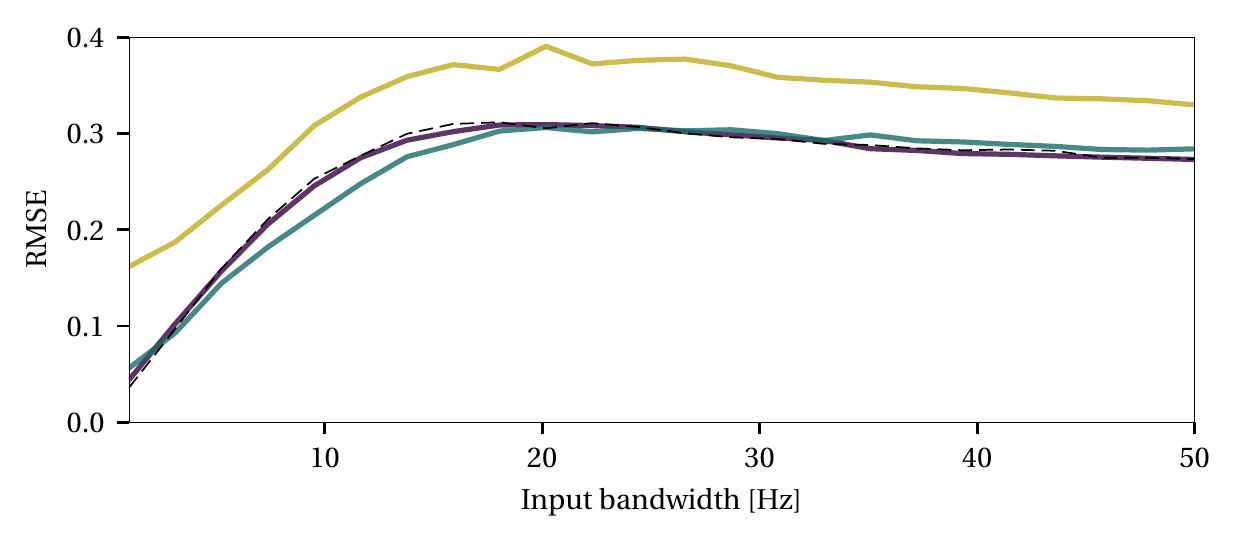}
		\caption{$n = 8$}
	\end{subfigure}
	\caption{Communication channel benchmark. A bandwidth limited white noise signal is fed into a series of neuron ensembles (where the number of ensembles $n$ is the \enquote{communication channel length}), with a synaptic time constant of $\SI{5}{\milli\second}$ for all connections. The graphs show the RMSE between input and output signal averaged over 16 runs. The dashed line depicts the baseline error obtained when filtering the input $n$ times with the synaptic filter.}
	\label{fig:benchmark_com_channel}
\end{figure}

\paragraph{Communication channel experiments}
The communication channel experiment consists of a chain of $\ell$ neuron ensembles with one hundred neurons each. All connections compute the identity function. The input to the system is a bandwidth limited white noise signal. As a benchmark measure, we compare the decoded output of the last ensemble in the chain to the input value. The experiment is executed for standard LIF neurons, LIF neurons with conductance-based synapses and eliminated $J^\mathrm{bias}$ (see \cref{sec:bias}), as well as LIF neurons with conductance-based synapses and active $J^\mathrm{bias}$. Results are shown in \cref{fig:benchmark_com_channel}.

As is clearly visible, there is no significant difference in error for the three neuron types for short communication channels. Notably however, across all experiments, current-based synapses are always best for inputs with low bandwidth, and using $J^\mathrm{bias}$ in conjunction with conductance-based synapses increases the error. Apart from this, conductance-based synapses are rather competitive compared to current-based synapses, and as the bandwidth increases, the error is mostly dominated by the baseline error obtained when filtering the input signal with the synaptic filter. Interestingly, current-based synapses are only better in high bandwidth scenarios when increasing the communication channel length to eight. At least the error caused by $J^\mathrm{bias}$ can be explained with the observations laid out in \cref{sec:ev}. Specifically, the average membrane potential deviates from the estimated values for a large input spike rate range.

\begin{figure}[t!]
	\centering
	\begin{subfigure}[t]{\linewidth}
		\centering
		\includegraphics{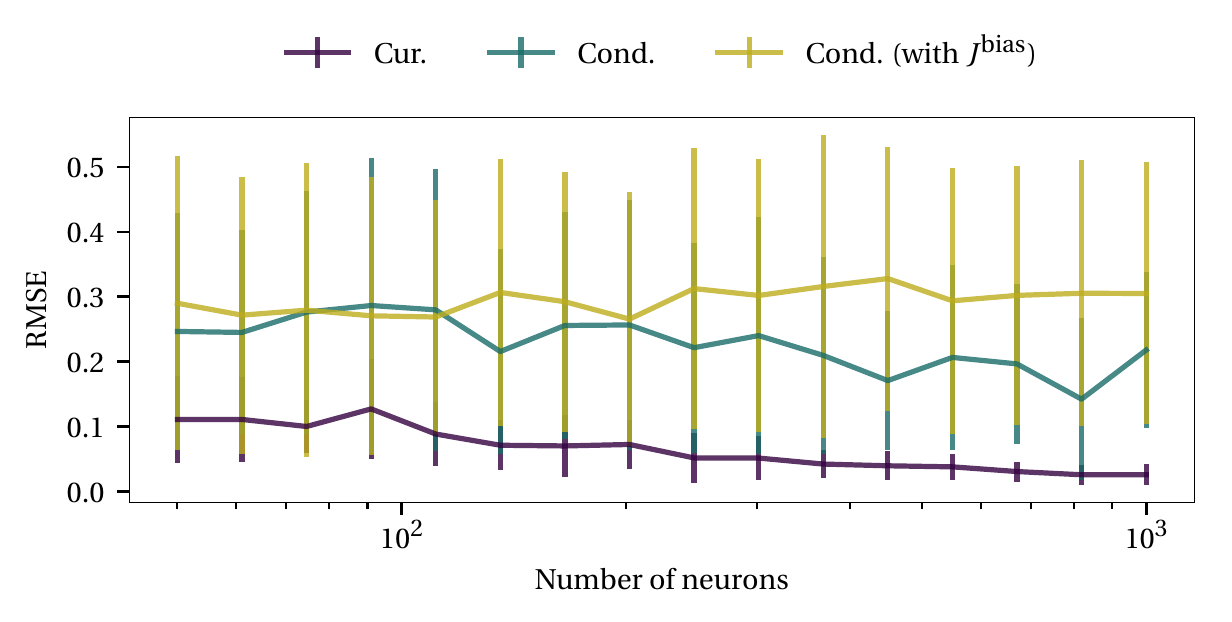}
		\caption{Integration error}
		\label{fig:benchmark_integrator_a}
	\end{subfigure}
	\begin{subfigure}[t]{\linewidth}
		\centering
		\includegraphics{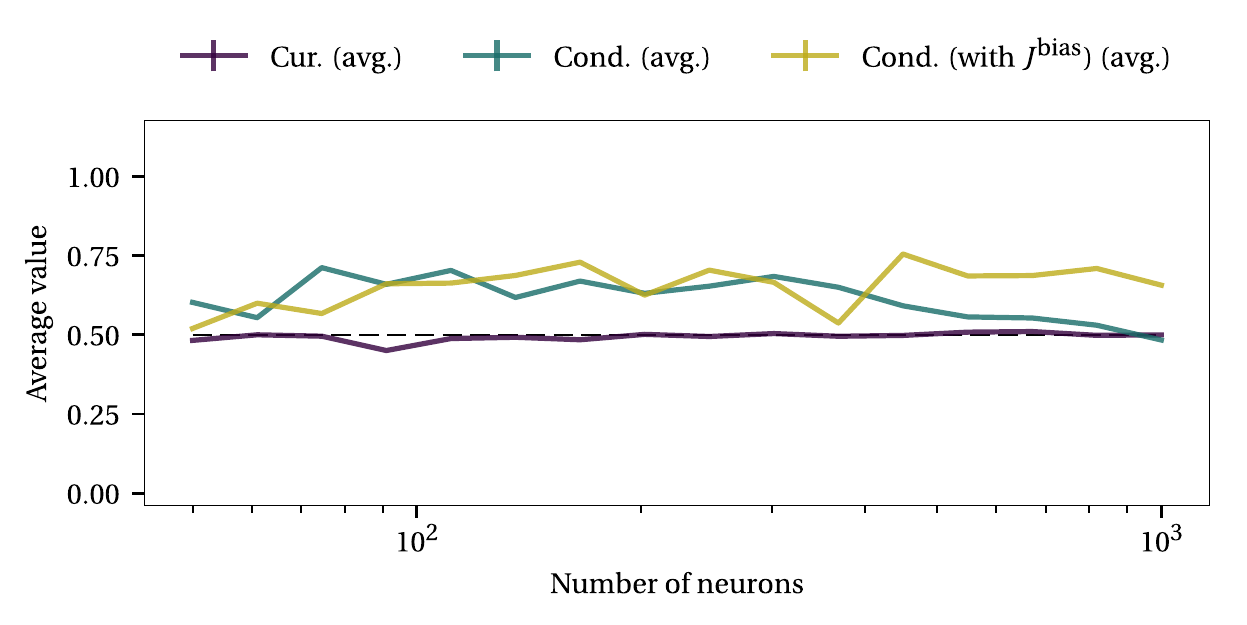}
		\caption{Average integrator value}
		\label{fig:benchmark_integrator_b}
	\end{subfigure}
	\caption{Results for the integrator benchmark. \textbf{(a)} Shows the average integrator representation error over 24 runs in which the integrator is driven to values in the interval $[0, 1]$. Error bars correspond to the standard deviation. \textbf{(b)} Shows the average value that the integrator converges to over all 24 runs. The dashed line depicts the expected value of $0.5$.}
	\label{fig:benchmark_integrator}
\end{figure}

\paragraph{Integrator experiments}
In the integrator experiment, a standard NEF integrator with $\SI{100}{\milli\second}$ synaptic time constant for the recurrent connection is tested. A value $x \in [0, 1]$ is fed into the integrator for one second, theoretically causing the integrator to converge to $x$. The simulation is continued for nine seconds and the difference between $x$ and the actual value is recorded. \Cref{fig:benchmark_integrator} shows the results over varying integrator neuron counts. For standard current-based synapses, the error is relatively small and converges to zero with increasing neuron count. The integrator with conductance-based synapses exhibits much larger errors. The error does not significantly decrease with increasing neuron count and is significantly higher when not eliminating $J^\mathrm{bias}$. When recording the integrator value trace over time, the integrator often converges to significantly larger than desired values. This is shown in \cref{fig:benchmark_integrator_b}, which depicts the average final integrator value. While the integrator with current-based synapses on average converges to the value $0.5$, as expected when sampling uniformly from the interval $[0, 1]$, the integrator with conductance-based synapses converges to significantly larger values.

\begin{figure}[t!]
	\centering
	\begin{subfigure}[t]{\linewidth}
		\centering
		\includegraphics{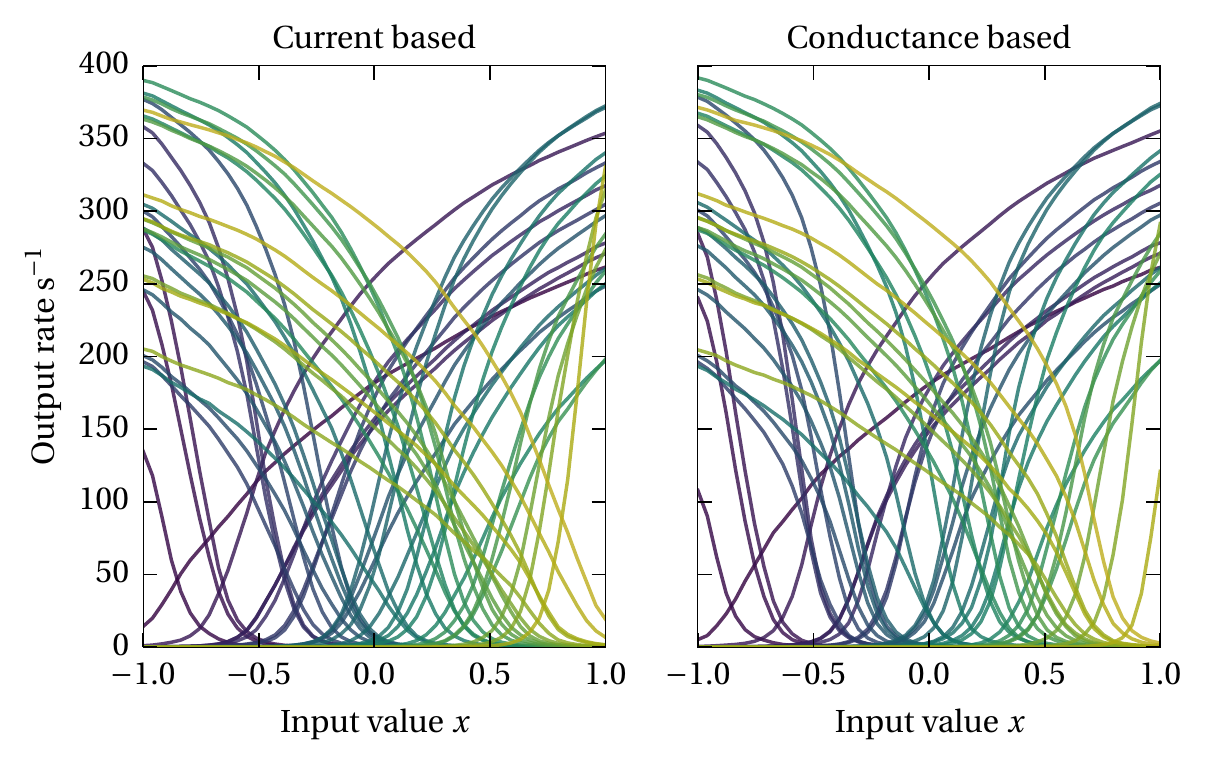}
		\caption{Tuning curves}
		\label{fig:tuning_curves_empirical}
	\end{subfigure}
	\begin{subfigure}[t]{\linewidth}
		\centering
		\includegraphics{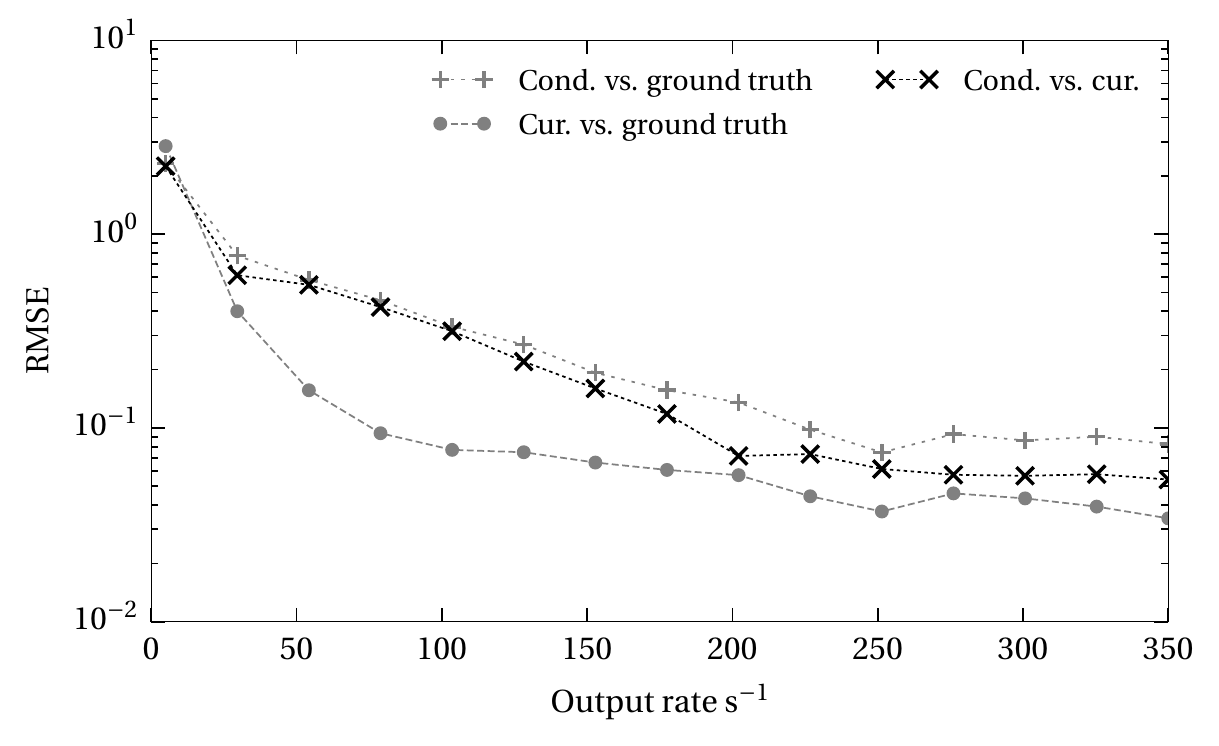}
		\caption{Error over output rate}
		\label{fig:tuning_curves_error}
	\end{subfigure}
	\caption{Tuning curve experiment results. (a) Empirically measured tuning curves for both current and conductance-based synapses. (b) RMSE between the
	tuning curves for varying output rate regimes. Ground truth refers to the non-spiking LIF firing rate model. In the \enquote{cond.~vs.~cur.} graph, the empirical measurements for the current-based model are used as a reference.}
\end{figure}

\paragraph{Empirical tuning curve measurment}
\Cref{fig:tuning_curves_empirical} shows empirically measured tuning curves for a LIF population with either current or conductance-based synapses. Tuning curves are measured by feeding a bandwidth limited (\SI{5}{\hertz}) white noise signal $x(t)$ into a standard current-based LIF ensemble with $1000$ neurons connected to the target population with $100$ neurons. The neural activities $a(x(t))$ are recorded, and the inverse mapping $x(a)$ is calculated. The measured tuning curves are similar, however, as shown in \cref{fig:tuning_curves_error}, they mainly differ for low output rates---the output rates of conductance-based neurons systematically are too small.
When simulating dynamical systems such as an integrator, the imbalance caused be the later spike onset (which prevents both excitation and inhibition) may be a reason for the systematic tendency to over-estimate the integral of the signal.

\paragraph{Summary}

In the feed-forward case, the simplistic transformation presented in \cref{sec:nef_cond} allows neuron ensembles with conductance-based synapses to function as well as their current-based counterparts, given that $J^\mathrm{bias}$ is eliminated from the neuron model (as outlined in \cref{sec:bias}). For mostly unknown reasons, the performance of the conductance-based synapses breaks down once recurrent networks are implemented. This may be caused by delayed spike onset in the tuning curves not accounted for when computing decoders.

\section{Conclusion}
\label{sec:conclusion}

In this report, we described both the NEF and LIF neurons with con\-duc\-tance-based synapses. We presented a simple transformation translating existing current-based NEF networks into their conductance-based counterparts. A sequence of benchmark experiments shows that this transformation yields relatively little error for feed-forward networks. However, when implementing dynamics such as integrators there are systematic deviations from the desired behaviour. Experiments show that the neuron tuning curves differ significantly between neurons with current-based synapses and the transformed conductance-based version. They exhibit a delayed spike onset, which is a potential explanation for the observed deviations in recurrent networks.

Unfortunately, this problem can likely not be solved with a simple transformation of pre-computed weight matrices. Instead, the two-dimensional neuron response curve of a neuron with excitatory and inhibitory conductance-based synapses must be taken into account. While the neuron response curve can be described analytically for constant, noiseless input, the non-negativity of weight matrices $W^+$ and $W^-$ mandates an iterative optimization scheme. Preliminary research indicates that these methods are feasible, yet computationally more expensive than standard NEF solvers, especially because the entire weight matrix $W$ is optimized and not the significantly smaller and reusable decoder $D$.

\printbibliography

\pagebreak

\appendix

\section{Expected value of a filtered, regularly spaced spike train}
\label{app:avg_filter_value}

\newcommand{\isi}{\mathit{\Delta t}}

\begin{figure}
	\includegraphics{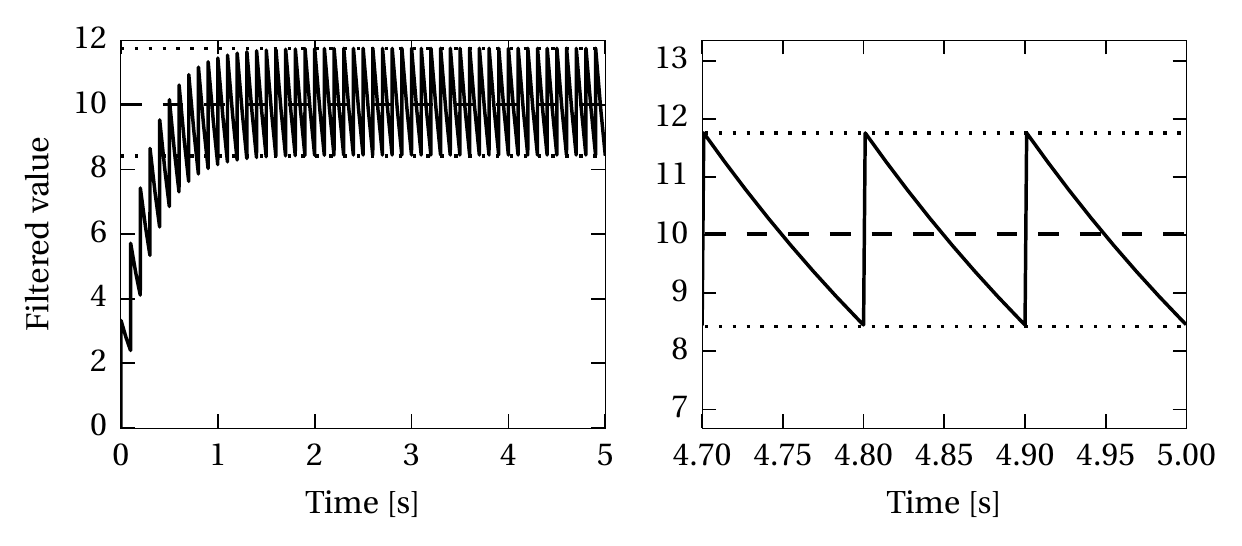}
	\caption{Example of a low-pass filtered regular spike train. In this example, individual input spikes are $\isi = 0.1$ apart. The filter time constant is $\tau = 0.3$. The right graph is a magnification of the left one. Dotted lines correspond to $\hat h$ and $\hat h - \tau^{-1}$, the dashed line to the final average $\isi^{-1} = 10$.}
	\label{fig:average_g_sample}
\end{figure}

\paragraph{Claim}
Given an infinite spike train with regular spacing $a(t) = \sum_{i = 0}^\infty \delta(t - \isi \cdot i)$ and a normalized exponential low-pass filter $h(t) = \frac{1}{\tau} \cdot e^{-t/\tau}$ it holds
\begin{align*}
    \langle h \ast a \rangle = \lim_{t \to \infty} \int_0^t \frac{(h \ast a)(t')}{t} \,\mathrm{d}t' = \frac{1}\isi  \,.
\end{align*}

\paragraph{Proof}
This directly follows from linear systems theory and the unit-gain of the exponential low-pass. Alternatively, the supremum $\hat h$ of $h \ast a$ can be expressed in terms of a geometric series
\begin{align*}
        \hat h = \sup(h \ast a) =
		\frac{1}{\tau} \cdot
		\sum_{\ell = 0}^{\infty}
			e^{-\frac{\isi \cdot \ell}{\tau}} = \frac{1}{\tau} \cdot \frac{e^\frac{\isi}{\tau}}{e^\frac{\isi}{\tau} - 1} \,.
\end{align*}
Correspondingly, and as illustrated in \cref{fig:average_g_sample}, after a finite amount of time the filtered spike train approximately oscillates between $\hat h$ and $\hat h - \tau^{-1}$ every $\isi$. The function between these two points is an infinite sum of exponentials, which itself is an exponential $g(t) = a \cdot e^{-bt}$, with $g(0) \overset{!}= \hat h$ and $g(\isi) \overset{!}= \hat h - \tau^{-1}$. Solving for $a$, $b$ yields $a = \hat h$ and $b = \tau^{-1}$. The average of $g(t)$ over $[0, \isi)$ is
\begin{align*}
	\langle h \ast a \rangle = \langle g(t) \rangle
	&= \frac{1}{\isi} \cdot \int_{0}^{\isi} g(t) \,\mathrm{d}t
	 = \frac{1}{\isi \cdot \tau} \cdot \frac{e^\frac{\isi}{\tau}}{e^\frac{\isi}{\tau} - 1} \cdot \int_{0}^{\isi} e^{-\frac{t}{\tau}} \,\mathrm{d}t \\
	&= \frac{1}{\isi} \cdot \frac{e^\frac{\isi}{\tau}}{e^\frac{\isi}{\tau} - 1} \cdot \left(1 - e^{-\frac{\isi}{\tau}}\right)
	= \frac{1}{\isi} \cdot \frac{e^\frac{\isi}{\tau}}{e^\frac{\isi}{\tau} - 1} \cdot \frac{e^{\frac{\isi}{\tau}} - 1}{e^{\frac{\isi}{\tau}}}
	=\frac{1}{\isi} \,. \quad \Box
\end{align*}

\end{document}